\documentclass[aip,rsi, amsmath,amssymb,reprint]{revtex4-1}
\pdfoutput=1
\usepackage[colorlinks=true,linkcolor=blue,citecolor=blue,allcolors=blue]{hyperref}
\usepackage[dvipsnames]{xcolor}
\usepackage{gensymb}
\usepackage{graphicx}
\usepackage{dcolumn}
\usepackage{multirow}
\usepackage{bm}
\usepackage[utf8]{inputenc}

\begin{document}

\title{Shear thinning and thickening in spherical nanoparticle dispersions}

\author{E. K{\"u}{\c c}{\"u}ks{\"o}nmez}
\email{ekinkucuksonmez@itu.edu.tr}

\author{J. Servantie}
\email{cservantie@itu.edu.tr}

\affiliation{Department of Physics, Istanbul Technical University, Maslak 34469, Istanbul, Turkey}

\date{\today}

\begin{abstract}
  We present a molecular dynamics study of the flow of rigid spherical
  nanoparticles in a simple fluid. We evaluate the viscosity of the
  dispersion as a function of shear rate and nanoparticle volume
  fraction. We observe shear thinning behavior at low volume
  fractions, as the shear rate increases, the shear forces overcome
  the brownian forces, resulting in more frequent and more violent collisions
  between the nanoparticles. This in turn results in more
  dissipation. We show that in order to be in the shear thinning regime
  the nanoparticle have to order themselves into layers
  longitudinal to the flow to minimize the collisions. As the
  nanoparticle volume fraction increases there is less room to form
  the ordered layers, consequently as the shear rate increases the
  nanoparticles collide more which results in turn in shear
  thickening. Most interestingly, we show that at intermediate volume
  fractions the system exhibits metastability, with successions of
  ordered and disordered states along the same trajectory. Our results
  suggest that for nanoparticles in a simple fluid the hydro-clustering
  phenomenon is not present, instead the order-disorder transition is the leading mechanism for the
  transition from shear thinning to shear thickening. 
\end{abstract}

\maketitle

\section{Introduction}

Viscosity is one of the fundamental physical property of fluids. It defines a
fluid's resistance to flow. The phenomenological law relating the shear stress to the
shear rate is Newton's law of viscosity \cite{JKatz},
\begin{equation}
\tau =\eta(\dot{\gamma})\ \dot{\gamma}, 
\end{equation}
where $\tau$ is the shear stress, $\dot{\gamma}=\partial v_x/\partial z$ the shear rate, and $\eta$
the shear viscosity. Shear viscosity quantifies the rate of momentum transfer per unit area between
two adjacent layers of fluid. A large viscosity results in higher momentum transfer, at the
limit $\eta\to\infty$ the system behaves like a solid and all the momentum is transferred. For the
so-called Newtonian fluids, the shear viscosity is independent of the
shear rate. Most fluids are Newtonian for small shear rates, the so-called newtonian plateau. However, many fluids show
non-Newtonian behavior at higher shear rates, usually one observes a decreasing viscosity with
increasing shear rate. The fluid flows easier as it becomes faster. This phenomenon is called
shear-thinning and is observed in fluids such as polymer
melts\cite{Doi,Chang}, colloid or non-colloids dispersions \cite{Jaeger14,bounoua16} and even nano-confined water \cite{Kapoor14}.

On the other hand, some fluids
exhibit the opposite behaviour, after a critical shear rate, $\dot{\gamma}_c$, flow becomes
more difficult, viscosity increases, this is called shear-thickening.  Shear-thickening is
generally observed in suspensions and colloidal dispersions \cite{Barnes89,Jaeger14,Crawford2013}. At high
volume fractions of colloids and high shear rates, shear-thickening can lead to a diverging
viscosity. This was observed in in early experiments with suspensions
of spherical particles  by Hoffman\cite{Hoffman72,Hoffman74,hoffman1991}, spherical colloid
  dispersions by Bender and Wagner \cite{wagner96} and recently in cornstarch suspensions by Madraki
  {\it et al.} \cite{Madraki2017}.

Shear-thickening can have negative impacts on engineering and industrial applications of materials
such as cement or coating dyes \cite{Cement,Dye}. However, it can also be a useful property, for example
for the fabrication of soft armors \cite{Lee2003,Gong2014} or sound insulation \cite{Li2014}. In either
case, it is important to have an understanding of the microscopic dynamics leading
to shear-thickening. Shear-thickening depends on several parameters. Primarily the volume
fraction of solid particles, $\phi$. Indeed, experiments \cite{Jiang2014,Chow1995,Fall2015,Cwalina2016} and
simulations \cite{foss_brady_2000,Seto2013,Mari2015,Pednekar2017} show that shear-thickening occurs
only after a minimum volume fraction of particle is reached in the fluid. The size of the particles
is an other important parameter. It affects the critical shear rate, the larger the particles are
the smaller is $\dot{\gamma}_c$, thus the onset of thickening is at lower values
of the shear rate \cite{Wagner2001,Wang2017}. The interaction between the fluid and solid
particles is also of importance. Indeed, if the particle-fluid interaction is too repulsive --or the
particle-particle interaction too attractive-- the solid particles will tend to aggregate, consequently
the fluid will loose its characterization of suspension or dispersion, and become unstable. The rheology
of aggregating fluids is another area of research
\cite{Wolthers}.

The shear thickening phenomenon is divided into two classes,
discontinuous shear thickening (DST) for which the viscosity increases of
several order of magnitudes. DST occurs over a critical volume
fraction of particles \cite{Seto2013}. DST is now relatively well
understood, it is caused by frictional contact between the suspended
particles \cite{Peters2016}  which can eventually lead to a jamming 
state \cite{Fall2008,Jaeger10}. The second mechanism is called
continuous shear thickening (CST) where viscosity increases slowly. Two alternative mechanisms for CST were
proposed. First the so-called order-disorder transition (ODT) suggested
by Hoffmann \cite{Hoffman72,Hoffman98} . The experiments on
concentrated colloidal suspensions suggested that shear-thickening occurs when
the suspension has a transition from an ordered micro-structure to a
disordered one. Later experiments by Ackerson and Pusey
\cite{Ackerson1988} Yan {\it et al.} \cite{Yan1994} also observed the
formation of ordered layers or strings of colloids. 
At low shear rates the suspended particles
flow in ordered layers while at high shear rates their flow becomes
disordered. This results in increased 
collisions and consequently increased frictional interactions and eventually shear thickening. 

The second mechanism is the so-called hydro-clustering phenomenon. Hydro-clusters were first
observed by Brady and Bossis \cite{brady_bossis1985} in Stokesian dynamics simulations, later experimental evidences were
observed by Wagner and coworkers\cite{maranzano2002,Kalman2009} and
Cheng {\it et al.}\cite{Cheng2011}. The results suggests that for large shear rates, the forces due to the flow overwhelms
the repulsive forces between the solid particles. This results in the formation of transient clusters. The lubrication
forces acting on the interstitial fluid causes an increased dissipation and consequently larger viscosity. As the shear
rate increases the size of the clusters increase, resulting in shear thickening.

The aim of this paper is to elucidate which mechanism is relevant for
dispersions of spherical nanoparticles. To achieve this we model a
suspension of nanoparticles with coarsed-grained molecular
dynamics simulations.

The manuscript is organized as follows: In Sec.~\ref{sec2} we describe our
simulation model and technique.  Then, in Sec.~\ref{sec3}, we compute the viscosity as a
function of shear rate for different volume fractions of nanoparticles. We relate the thinning
or thickening behavior of the fluid to the microscopic structure of
the fluid where we show that at large volume fractions thinning occurs when the nanoparticles
can order themselves in order to minimize the number of collisions. The manuscript
closes with a brief discussion in Sec. \ref{sec5}

\section{\label{sec2} The Model}

In this work we are interested in the universal properties leading to thinning or thickening in
suspensions. We thus construct a coarse-grained model. The particles of the base fluid interact
through a Lennard-Jones (LJ) potential, 
\begin{equation}
	V_{LJ}=\left \lbrace \begin{array}{ccc}
	4\epsilon\left[(\frac{\sigma}{r})^{12}-(\frac{\sigma}{r})^{6} \right] & {\rm for} &  r < r_c \\
0 &	{\rm for} & r \geq r_c 
	\end{array} \right.
\end{equation}
where the cutoff distance is chosen to be $r_c=2.5 \sigma$. The Lennard-Jones parameters are fixed to
unity, $\epsilon=1$ and $\sigma=1$. The mass of the particles is also fixed to unity $m=1$. A unit of time can thus be
expressed as $\tau=\sigma \sqrt{m/\epsilon}$. The nanoparticles are modeled as rigid molecules of spherical shape
with a radius of 1 $\sigma$. They consist of 100 atoms, which is enough to ensure fluid atoms can not
enter inside them. The interactions between the nanoparticles and the fluid, and between the nanoparticles
is modelled with a modified Lennard-Jones potential in order to control the hydrophobicity of the
nanoparticles,
\begin{equation}
  V_{ab}=\left \lbrace
  \begin{array}{ccc}
    4\epsilon\left[(\frac{\sigma}{r})^{12}-C_{ab}(\frac{\sigma}{r})^{6} \right] & {\rm for} &  r < r_c \\
    0 &	{\rm for} & r \geq r_c 
  \end{array} \right.
\end{equation}
$a,b$ indicates the type of atom, $f$ for fluid atoms and $n$ for atoms of nanoparticles. The
parameter $C_{ab}$ controls the strength of the attractive part.  $C_{nn}=0.2$ and $C_{nf}=0.5$ permits to
have a well dispersed nanofluid.  

\begin{figure}[h!]
\includegraphics[scale=0.45]{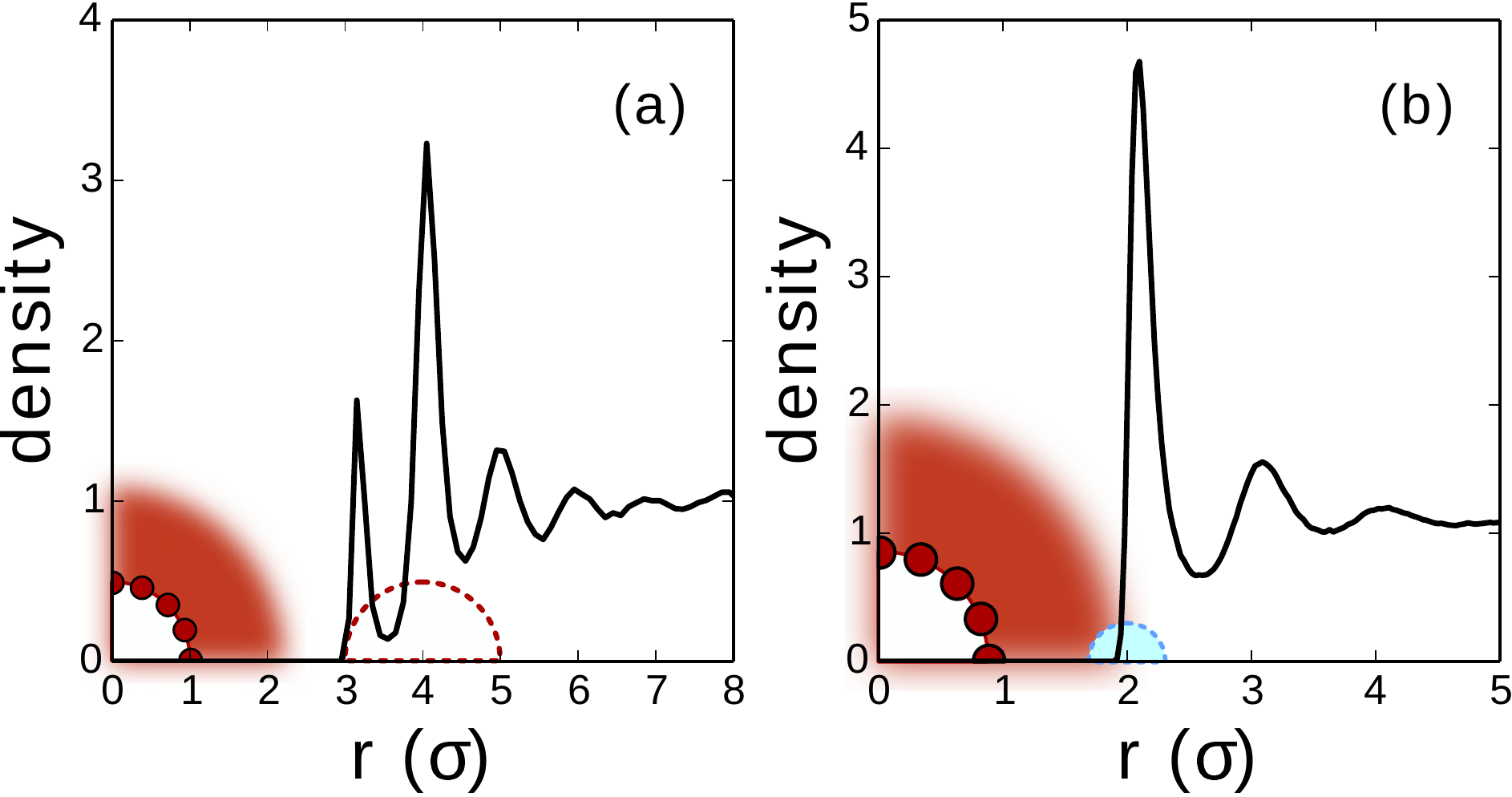}
\caption{\label{fig:pair} (a) Pair correlation function between nanoparticles, the dashed line represents a nanoparticle
  at the most probable separation, the blurred region corresponds to the effective size of a nanoparticle,
  approximately $2\ \sigma$. (b) Pair correlation between nanoparticles and fluid atoms, the dashed line represents a fluid atom
at its most probable distance to the nanoparticle.}
\end{figure}

\begin{figure}[h!]
\includegraphics[scale=0.33]{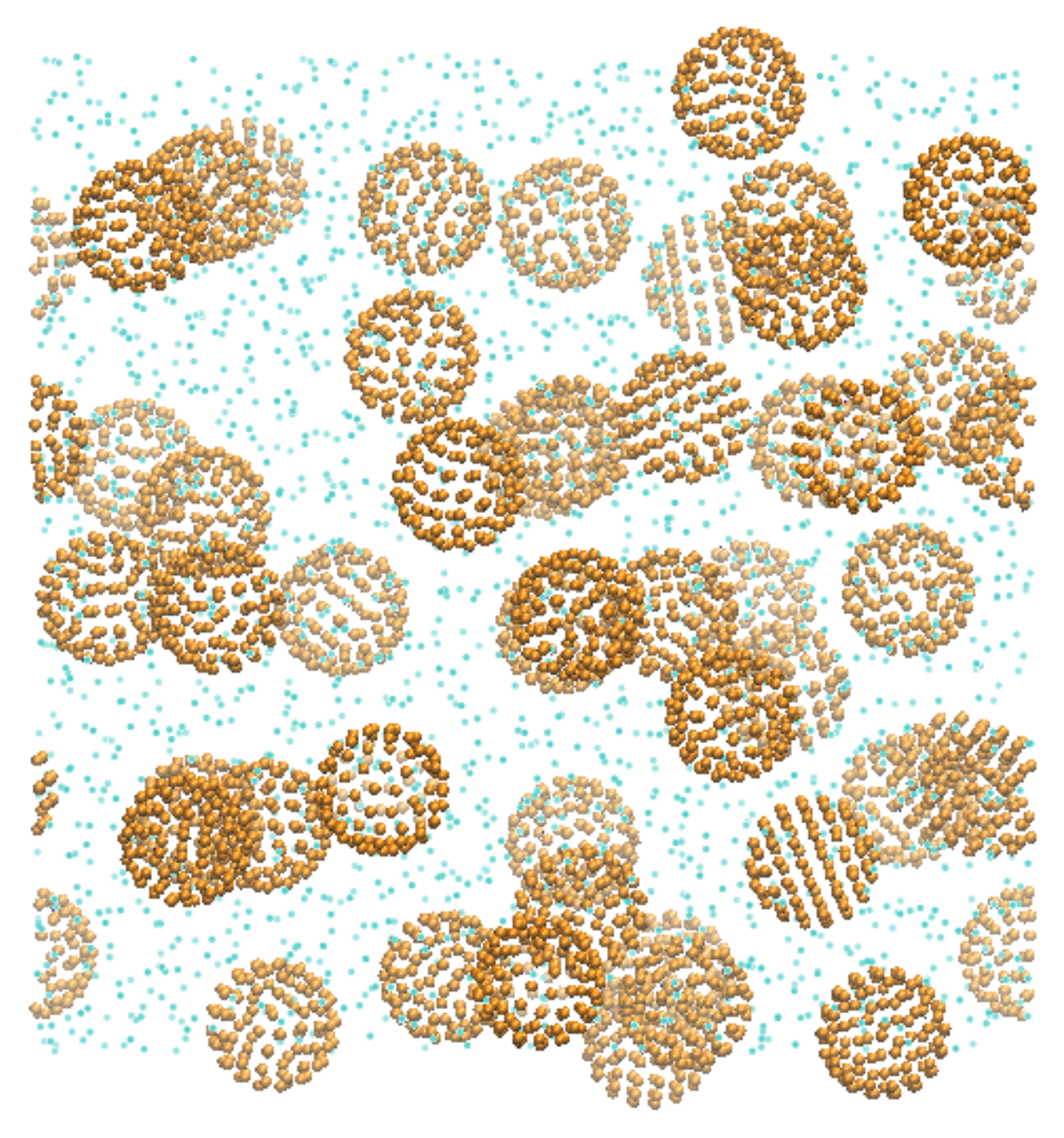}
\caption{\label{fig:rig} Snapshot of a suspension of spherical
  nanoparticles with a volume fraction 0.41 dispersed in a
  Lennard-Jones fluid.}
\end{figure}

The volume fraction of nanoparticles in the suspension can be written as, 
\begin{equation}
 \phi = N_p \frac{\frac{4}{3} \pi r_{p}^3 }{V}
\end{equation}
where $r_{p}$ is the effective radius of the nanoparticle, $N_p$ the
number of nanoparticles and $V$ is the volume of the simulation box. The effective radius of the
nanoparticle can be evaluated thanks to the radial pair correlation
function evaluated between the nanoparticles and nanoparticles, and
nanoparticles and the fluid as depicted in Fig. \ref{fig:pair}
Considering the radial distribution,  we define the effective radius
of the nanoparticle as $2\ \sigma$. The volume fractions are evaluated
with this value. Notice that according to Fig. \ref{fig:pair}(a) the most probable distance
of two nanoparticles is approximately $4\ \sigma$. Finally, we prepare cubical simulation boxes with $16\ \sigma$ of side length. The
nanoparticle volume fraction varies between $\phi=0$ to $\phi=0.53$. This
corresponds to a number of fluid atoms varying between 3648 and 1708,
and 0 to 65 nanoparticles.  We depict a typical system in Fig. \ref{fig:rig}.

\subsection{Equations of motion}

We use the rotation matrix algorithm to enforce the rigid body
motion. \cite{Dullweber,Rapaport} While the SHAKE or RATTLE \cite{SHAKE,RATTLE} algorithms
use constraints to enforce the rigid body motion, it is not the case for the rotation matrix
algorithm \cite{Sunarso2011,Akimov2011,Orsi2010}. One can thus derive a reversible integration
algorithm which in turn permits to do long simulations without any unphysical velocity
scalings.

The rotation matrix is the transformation that maps the moment of inertia tensor in the simulation box frame
of the molecule to the frame in which the moment of inertia tensor is
diagonal (principal axes frame). At every time step the rotation
matrix of each molecule is calculated, and the coordinates are
transformed into the principal axes frame, in which the equations of
motions are, 
\begin{eqnarray}
  \dot{\bm{r}}_{\rm CM}&=&  \frac{\bm{p}_{\rm CM}}{M} \\
  \dot{\bm{p}}_{\rm CM}&=&\sum \mathbf{F}_{i}  \\
  \dot{\bm{\theta}}&=&\bm{\omega} \\
  \bm{I} \dot{\bm{\omega}}&=&\sum \bm{\tau}_i 
\label{rigideqmotion}
\end{eqnarray}
where the moment of inertia matrix $\bm{I}$ is now diagonal and
constant. $M$ is the mass of the molecule, $\bm{r}_{\rm CM}$ and
$\bm{p}_{\rm CM}$  respectively, the
position of its center of mass and its momentum. Finally,
$\bm{\theta}$ and $\bm{\omega}$ are respectively the angular position and angular
velocity vectors of the molecule. The total force and
torque is calculated over all the interactions with fluid atoms and
atoms of other molecules. Angular accelerations, angular velocities, positions and
momenta of the center of mass are updated with a
velocity Verlet type scheme in the principal axes frame, then the coordinates are transformed
back to the simulation box frame. 

In order to compute the viscosity as a function of shear rate we impose a Couette flow on the
fluid. One can achieve a Couette flow either by adding a physical wall
to the system and give is a constant velocity a motion, or by changing
the boundary conditions for the bulk fluid 
in order to avoid surface effects. For
the latter one must use the Lee-Edwards or sliding brick periodic boundaries \cite{Lees_Edwards}. The
modification of the periodic boundaries leads to a change in the equations of motion. For point particles
one can use the isokinetic SLLOD algorithm \cite{Evans86,Evans95,EvansNESM}. The equations of motion for a
Couette flow in the $\hat{x}$ direction are written as, 
\begin{eqnarray}
\dot{\mathbf{r}}&=&\frac{\mathbf{p}}{m} + \dot{\gamma}z\ \hat{x} \label{SLLODrp1} \\
\dot{\mathbf{p}}&=&\mathbf{F} - \dot{\gamma}p_{z}\ \hat{x}-\zeta \mathbf{p}
\label{SLLODrp2}
\end{eqnarray}
where $\mathbf{F}$ is the total force exerted on the atom and $-\zeta
\mathbf{p}$ a frictional term to achieve a constant kinetic energy simulation.
For molecules one must use this algorithm with care. Indeed, the equations of motion in Eqs. \ref{SLLODrp1}
and \ref{SLLODrp2} can only be applied to a mono-atomic fluid. For molecules they have to be
modified to avoid the independent motion of atoms in a molecule. There are two possible approaches,
atomic SLLOD and molecular SLLOD \cite{Evans86,Evans95}. The atomic SLLOD equations of motion are the
same as the original except that a constraint is added to conserve the
molecular structure. On the other hand, in the case of molecular
SSLOD, the SSLOD algorithm is only applied to 
the center of mass of the molecule, hence
avoiding the use of constraints. Both algorithms give the same results as long as the shear rate is not
too large. The equations of motion for the center of mass of a
molecule are thus,
\begin{eqnarray}
  \dot{\mathbf{r}}_{{\rm CM}}&=&\frac{\mathbf{p}_{{\rm CM}}}{M} + \dot{\gamma}z_{CM} \ \hat{x} \\
  \dot{\mathbf{p}}_{{\rm CM}}&=& \mathbf{F}_{\rm CM}- \dot{\gamma}p_{{\rm CM}z}\ \hat{x} 
\label{mSLLODrp}
\end{eqnarray}
where $\mathbf{F}_{\rm CM}$ is the total force acting on the molecule. Remark that the frictional term $-\zeta \mathbf{p}$ is
not present for the molecules, since the number of fluid atoms is much
larger than the number of molecules, thermalization is quickly achieved only with the fluid atoms.

\subsection{Evaluation of the shear viscosity}

One can evaluate the shear viscosity with equilibrium molecular
dynamics (MD) thanks to the Green-Kubo relationship,
\begin{equation}
\eta = \beta V \int_0^{\infty} dt\ \langle P_{xz}(t)P_{xz}(0) \rangle
\end{equation} 
where $\beta$ is the inverse temperature, $V$ the volume of the
system, and $P_{xz}$ the $xz$ component of the pressure tensor. On the other hand, for non-equilibrium
molecular dynamics simulations (NEMD) one gets the viscosity directly
from Newton's law of viscosity,
\begin{equation}
 \eta = -\frac{\langle P_{xz} \rangle}{\dot{\gamma}}.
\end{equation}
Calculating the viscosity both from equilibrium MD and NEMD permits to validate the NEMD
algorithm.  The $xz$ component of the atomic pressure tensor is written as \cite{quarrie},
\begin{equation}
P^a_{xz}=\frac{1}{V}\left(\sum_i \frac{p_{xi} \ p_{zi}}{m_i} + \sum_{i<j} F_{xij}z_{ij} \right)
\label{pa}
\end{equation}
Where the momentum is the usual momentum for equilibrium simulations, or in case of the SLLOD equations
of motion they have to be taken as the peculiar momenta
\cite{Evans86,Evans95,EvansNESM} which correspond to the thermal
velocities, in other words independent from the shear applied to the
system. This expression can be used for the Lennard-Jones fluid, however, the atoms of molecules do not have
individual peculiar momenta because of the rigidity of molecules . For
molecules, one must consider the molecular pressure tensor
\cite{Evans86,Evans95}  which is determined in terms of the of
peculiar momenta of the center of mass of the molecules and
intermolecular forces acting on their center of mass, 
\begin{equation}
P^m_{xz}=\frac{1}{V}\left(\sum_i \frac{p_{{\rm CM},xi} \ p_{{\rm CM},
      zi}}{M_i} + \sum_{i<j} F_{{\rm CM},xij}z_{{\rm CM},ij} \right)
\label{pm}
\end{equation}
Remark that while the atomic pressure tensor has to be symmetric, it is not the case for the molecular
pressure tensor. The atomic and molecular pressure tensors are compared theoretically and
computationally in Refs. \cite{Allen84,Evans86,Evans95}. They are related to each other as  
\begin{equation}
P^a=P^m_{(S)}+\frac{1}{2}\ddot{\chi}
\end{equation}
where the subscript $(S)$ denotes the symmetrized molecular pressure tensor and $\chi$ is written as 
$$\chi=\sum_{i\alpha}m_{i\alpha}\delta\mathbf{r}_{i\alpha}\delta\mathbf{r}_{i\alpha}$$ 
where
$\delta\mathbf{r}_{i\alpha}=\mathbf{r}_{i\alpha}-\mathbf{r}_{{\rm CM},i}$. 
Where $i$ is the index of the molecule  and $\alpha$ is the index of
the atom in the molecule. $\langle \ddot{\chi} \rangle =0 $ for a system in a steady state. One can consequently
use the symmetrized molecular pressure tensor to evaluate the viscosity.

The viscosity of the dispersion is evaluated from the hybrid pressure
tensor which is the sum of the atomic pressure tensor in Eq. (\ref{pa}) and the symmetrized molecular pressure tensor
in Eq. (\ref{pm}) for our mixture of point-like atoms and spherical nanoparticles.

\section{\label{sec3}Results}

\subsection{Viscosity as a function of shear rate and volume fraction}

Starting from a pure Lennard Jones fluid, we evaluate the viscosity of
dispersions with different volume fractions $\phi$ as a function of
the shear rate, for values in the range $\dot{\gamma}=0\ 1/\tau$ to $\dot{\gamma}=2\ 1/\tau$. Unfortunately, we can not
compute higher values of the shear rate with the present
algorithm. Indeed, the molecular SLLOD algorithms breaks down at very high shear
rates\cite{Evans86,Evans95}.

\begin{figure}[h!]
\includegraphics[scale=0.45]{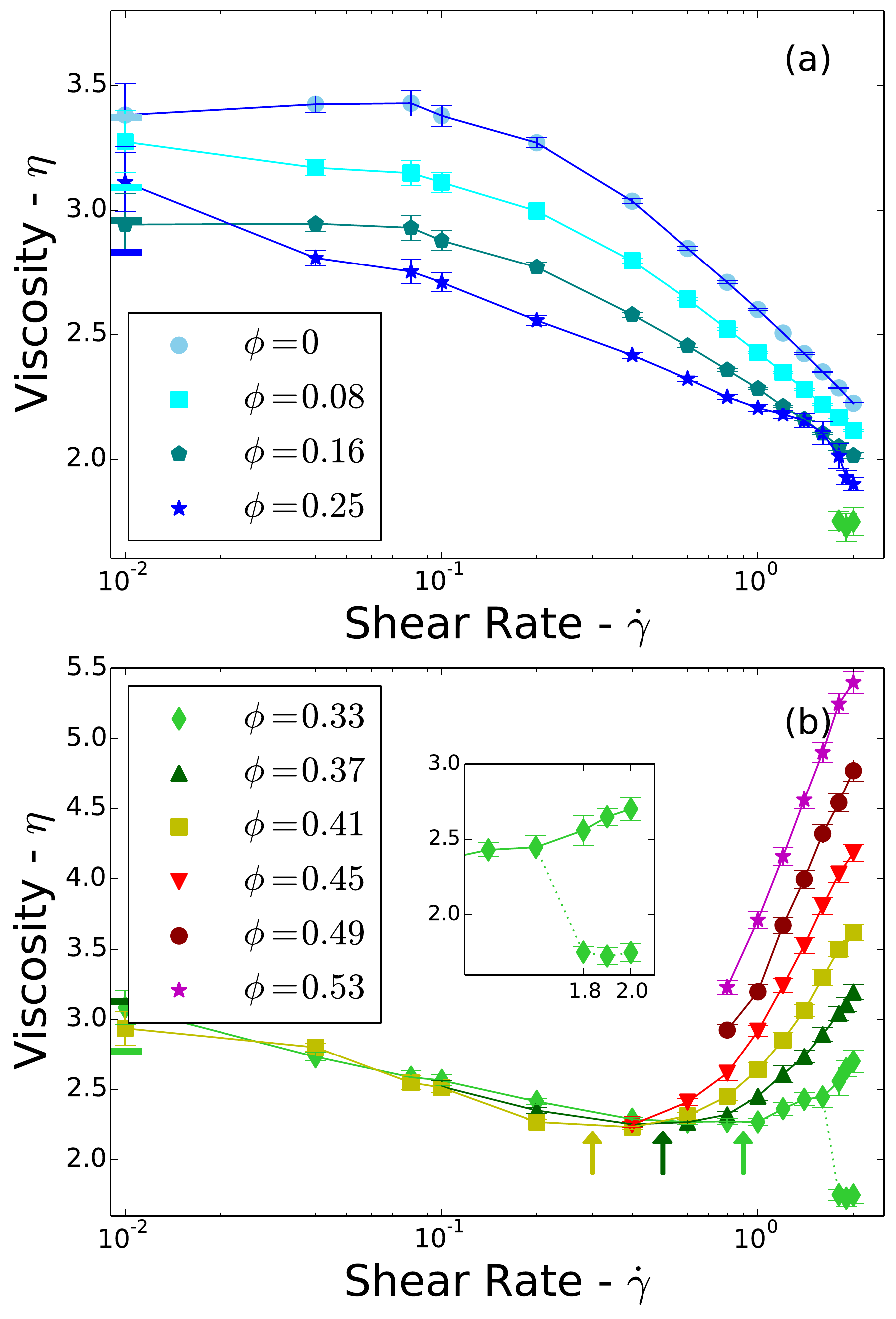}
\caption{\label{fig:shearvis} Viscosity  $\eta$ as a function of the shear
  rate $\dot{\gamma}$ for different volume fraction of
  nanoparticles. (a) small volume fractions exhibiting shear thinning,
  (b) large volume fractions exhibiting shear thickening. The
  arrows indicate the beginning of the shear-thickening regime. The horizontal lines at low
  shear rates correspond to the Green-Kubo results. The diamonds in
  (a) represent the viscosity for $\phi=0.33$ in the
thinning regime while the dashed line in (b) depicts the separation to
the thinning regime. The inset of (b) represents a close up view
on the bifurcation for the intermediate volume fraction $\phi=0.33$. The error
  bars correspond to the statistical error on an average with
  correlated sample. The solid lines serve as a guide to eye.}
\end{figure}

The simulations are performed with $10^6$ integration
step with a time step  $\Delta t=10^{-3} \tau$ after an equilibrium process
which ensures the system has reached a non-equilibrium steady state. For
shear rates smaller than $\dot{\gamma}=0.08\ 1/\tau$ 
the integrations run ten times longer to reduce the statistical
error. All the simulations are carried out at the temperature $k_B
T=1.2 \epsilon$.   We depict the results in Fig. \ref{fig:shearvis}.
We observe that for small shear rates, the NEMD results are in
agreement with the viscosity obtained from the equilibrium molecular
dynamics simulations thanks to the Green-Kubo relationship. As the
shear rate increases we observe two different behaviors, at low volume
fractions, namely about $\phi=0.3$ the fluid exhibits shear thinning
as depicted in Fig. \ref{fig:shearvis}(a). While for higher volume
fractions shear thickening occurs after a critical shear rate value
$\dot{\gamma}_c$ denoted by the arrows in Fig. \ref{fig:shearvis}(b). We observe
that the critical shear rate decreases as the volume fraction increases. For
volume fractions larger than $\phi=0.49$ the nanoparticle and liquid
mixture form a solid for small shear rates, the fluid is in a jammed
state. A steady state Couette flow can only be formed for large enough shear rates. Finally, the most interesting
behavior is for the intermediate volume fraction $\phi=0.33$, after the shear thinning regime a bifurcation occurs. Along the
same trajectory we observe sequences of thinning regime and thickening regime. Our simulation suggest the duration of each
sequence is random, however the thinning states appears to become longer with increasing shear rate. In order to elucidate this
metastable behaviour one has to study the microstructure of the fluid.

\subsection{Microscopic structure}

\begin{figure}[h!]
\includegraphics[scale=0.45]{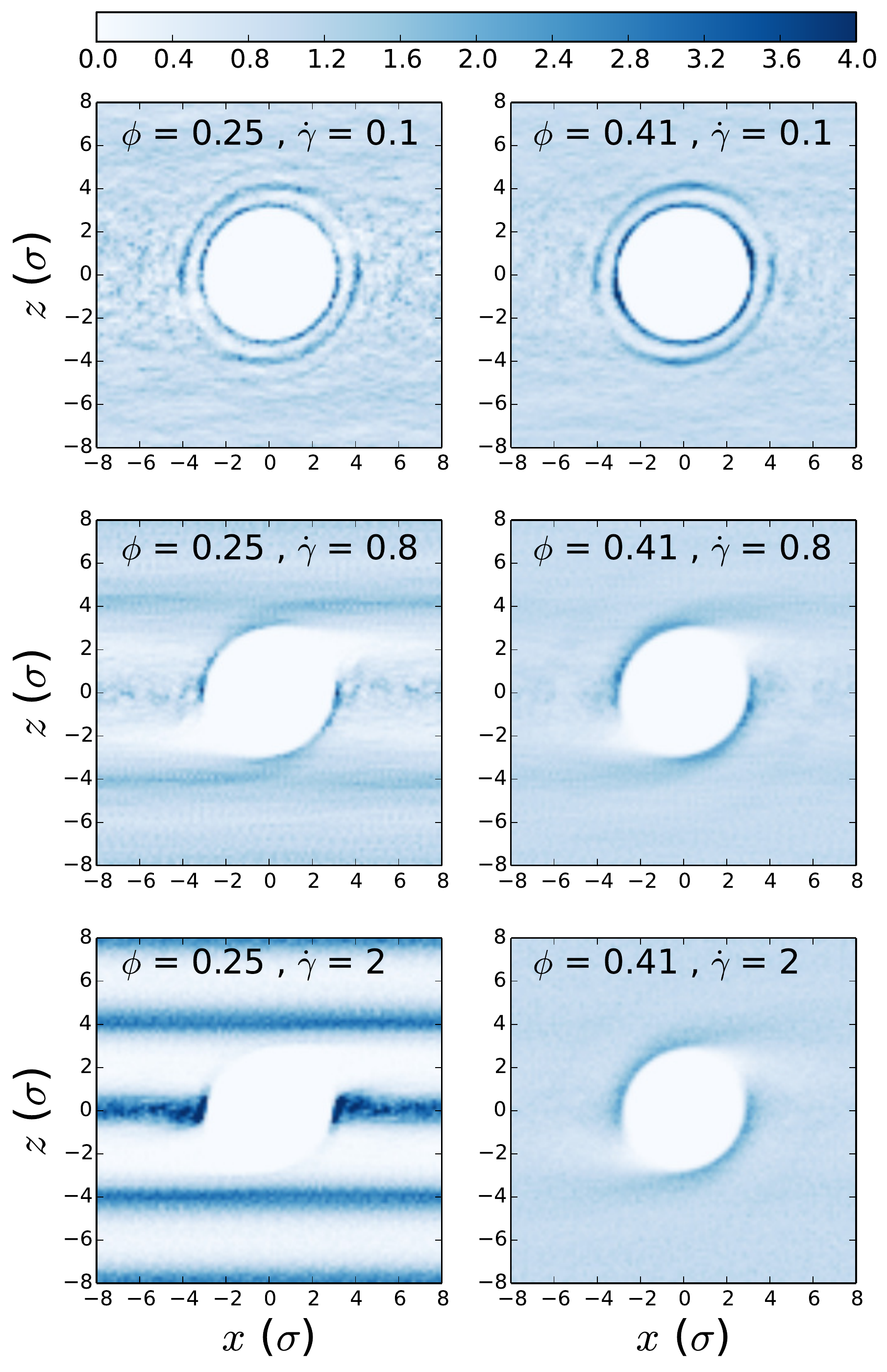}
\caption{\label{fig:2Dpair} Two dimensional pair correlations for
  volume fractions $\phi=0.25$ and $\phi=0.41$ for different shear rates. Darker regions correspond to regions
  of higher density.}
\end{figure}

In order to understand what happens at the microscopic scale to the dispersions in the different
viscosity regimes we evaluate the two-dimensional pair correlation
functions of the nanoparticles. The two-dimensional pair correlation
function is found by first  evaluating the pair correlation function in
three dimensions between nanoparticles. The result is then averaged over the $y$
direction and over all the nanoparticles. We depict in Fig. \ref{fig:2Dpair} the
two-dimensional pair correlation function for two different volume
fractions of nanoparticles, one which exhibits shear thinning,
$\phi=0.25$, and the other shear thickening, $\phi=0.45$. We evaluate
the correlation function for increasing values of the shear rate.
For small values of the shear rate both volume fractions exhibit shear
thinning, their pair correlation functions are also similar, in both
case the fluid is isotropic. However, as the shear rate increases, the pair correlation for $\phi=0.25$
shows a different behavior. One observes that layers appear and become
more apparent with increasing shear rate.  The distance between the
layers is approximately $4\ \sigma$, which corresponds to the most probable distance
between two nanoparticles as depicted in Fig. \ref{fig:pair}(b). As the
shear rate increases the nanoparticles follow trajectories in which
they avoid each other by forming a layered micro structure. It should be
noted that those layers are the result of statistical averages and are
not apparent when we analyze a single snap shot of the
simulation. Similar layers called sliding layers were observed experimentally
previously \cite{Yan1994,Lopez2015,Lee2018}. As the shear rate increases, the nanoparticles tend
to move in layers, hence avoiding collisions. This results in less energy dissipation,
and thus decreased viscosity. For $\phi=0.45$ no such layers are
formed, the nanoparticles can undergo violent collisions, which in
turn increases the viscosity of the dispersion.

\begin{figure}[h!]
\includegraphics[scale=0.45]{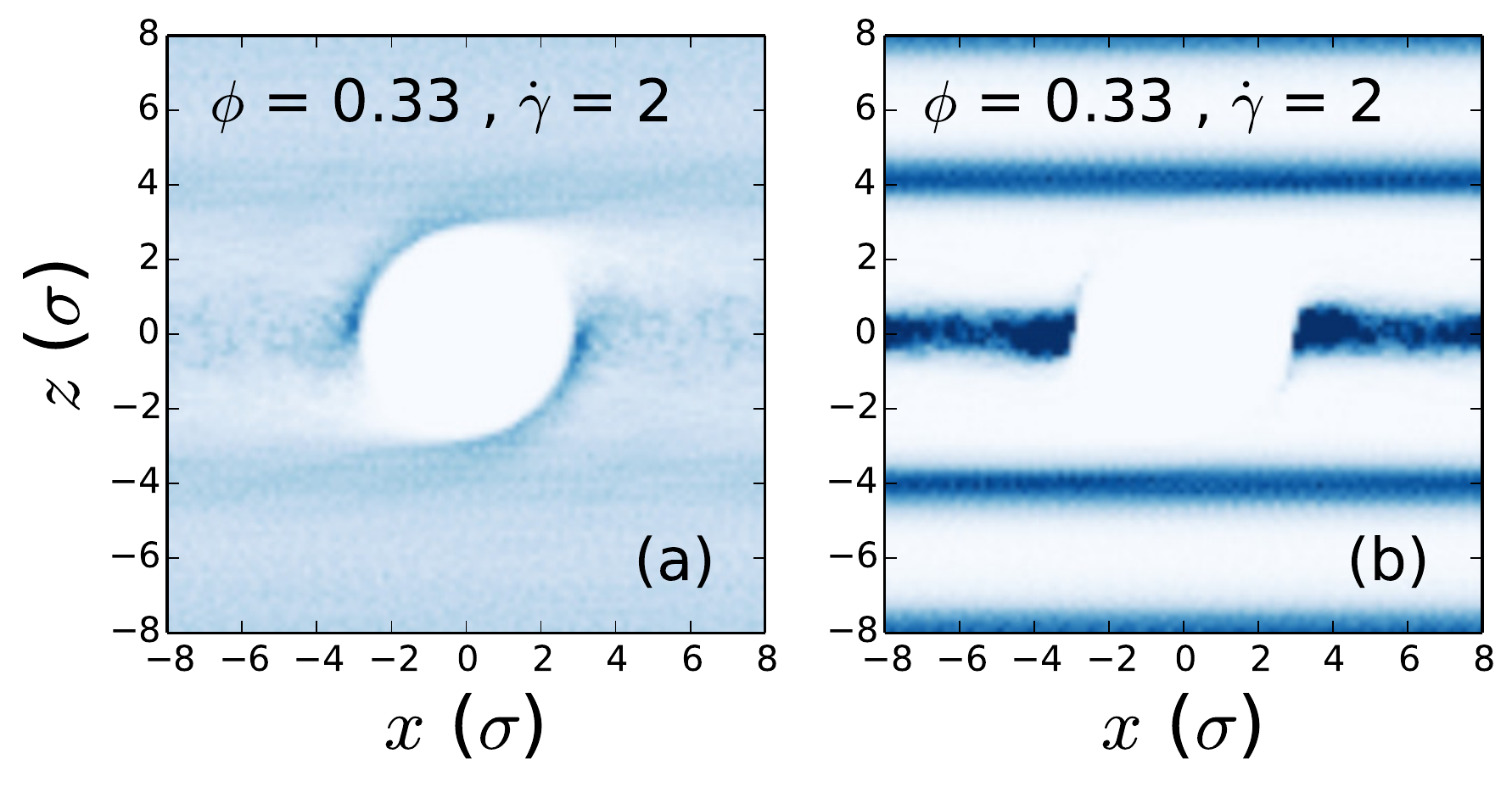}
\includegraphics[scale=0.45]{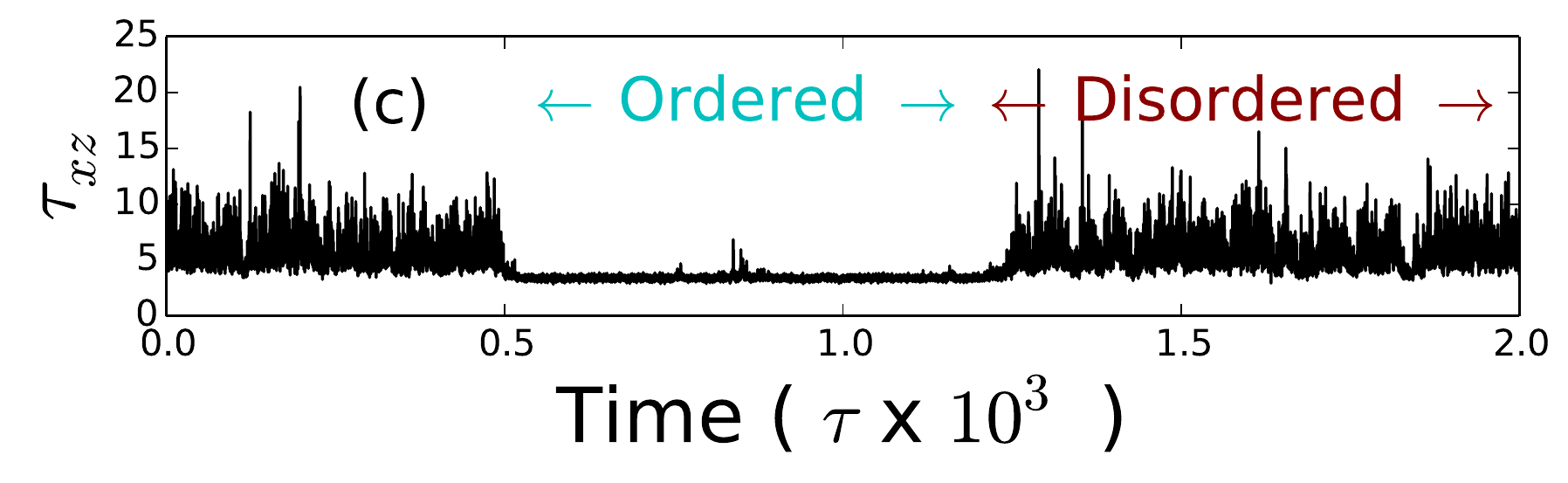}
\caption{\label{fig:2Dpair2} Two dimensional pair correlations for the intermediate volume fractions $\phi=0.33$ for the
  thickening regime (a) and thinning regime (b). (c) depicts the shear stress $\tau_{xz}$ as a function of time.}
\end{figure}

 We now focus on the intermediate volume fraction
$\phi=0.33$ which exhibits a metastable behavior at high enough shear rates. We depict in Fig. \ref{fig:2Dpair2} the two
dimensional pair correlation functions in the thinning and thickening regime and the shear stress as a
function of time. 
We observe that along the same trajectory the system
is first in a disordered state, in which the shear stress is high and
as a consequence the viscosity. After a while the nanoparticles order themselves for some finite duration thus decreasing
the viscosity significantly. This process repeats at random intervals,
hence suggesting the ordered state is metastable, large enough
fluctuations can disrupt the layers.

\begin{figure}[h!]
  \includegraphics[scale=0.45]{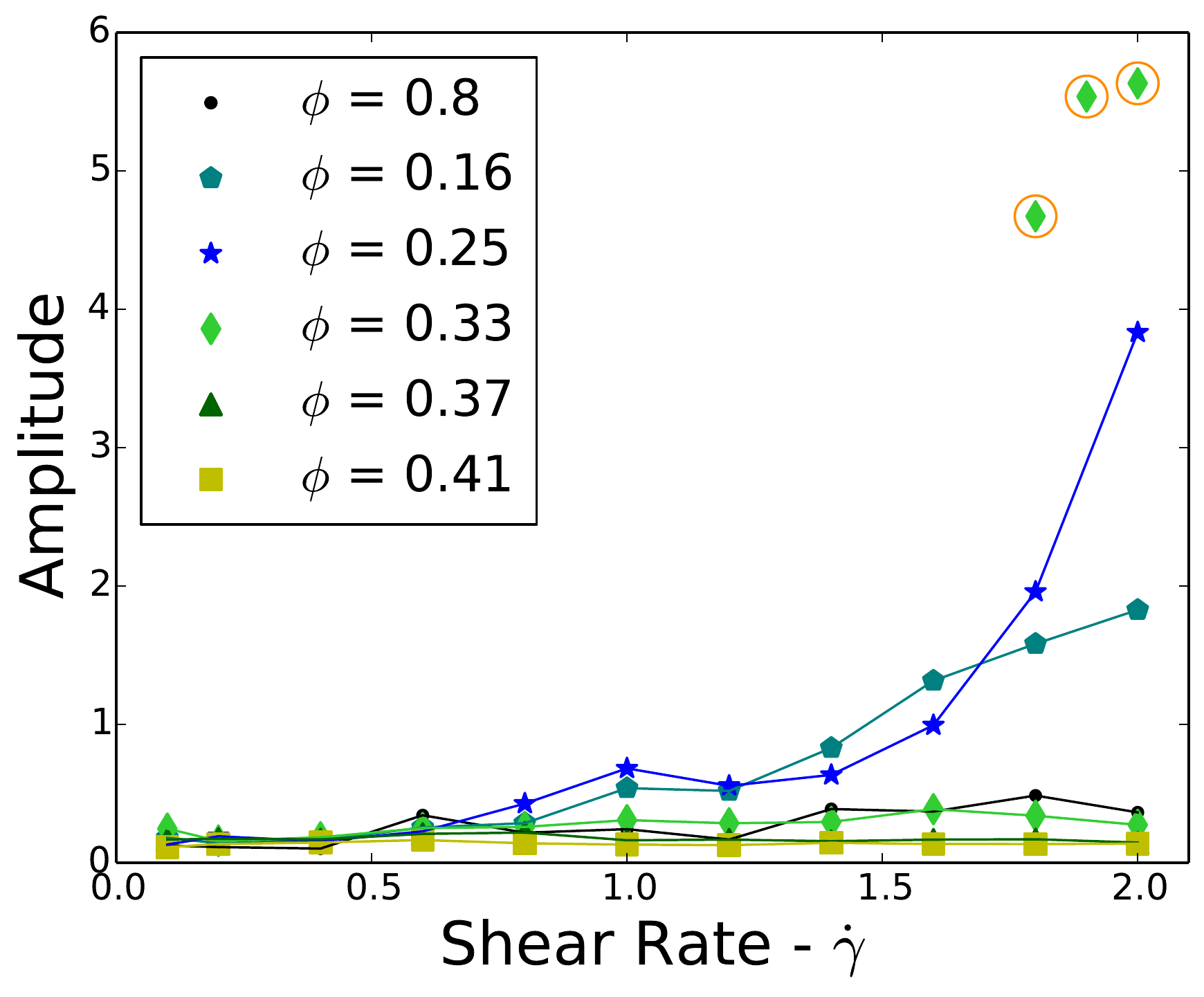}
  \caption{Amplitude of the power spectrum in the transverse direction to the flow at $z=4\ \sigma$ as a function
    of shear rate. The circles indicate the intermediate volume fraction $\phi=0.33$ in the thinning regime.}
  \label{fig:maxpower}
\end{figure}

In order to quantify the layering inside the fluid we
evaluate the power spectrum in the $z$ direction and depict its value for
$z=4\ \sigma$ as a function of shear rate in Fig. \ref{fig:maxpower}.

We see that for systems exhibiting shear thinning, the amplitude of the mode corresponding to the layer
width increases very fast with the shear rate while systems undergoing shear thickening remain
isotropic. Remark that the ordering increases continuously, we can not define a specific shear rate at which
ordering starts. In general, as the volume fraction increases the layering becomes more pronounced, this is due to the
fact that there are more nanoparticles in the layers, and thus a larger density.
This behavior is typical of the order-disorder
transition phenomenon proposed by Hoffman \cite{hoffman1991,Hoffman98}, as the shear rate
changes micro structures are formed by the nanoparticles to avoid the increase in internal stress and as a
consequence shear thickening. We remark that at very low volume
fractions the nanoparticles have very few interactions, as a
consequence they do not need to form sliding layers to decrease the
internal stress. The layers occur at high volume fractions and large
enough shear rates.   
 
On the other hand,  our results do not suggest hydro-clusters are formed in the shear thickening
regime. Instead the nanoparticles remain well dispersed. We believe that for nanoparticles the absence of surface
roughness, and thus macroscopic frictional effects prevents the formation of transient
clusters. It is interesting to see that for nanoparticles, the shear-thickening phenomenon does not require hydro-clusters to
form as suggested by several authors \cite{Pan,Guy,Mari,Heussinger,Seto,Denn,Fernandez2013}.

\subsection{Nanoparticle collisions}

\begin{figure}[h!]
\includegraphics[scale=0.45]{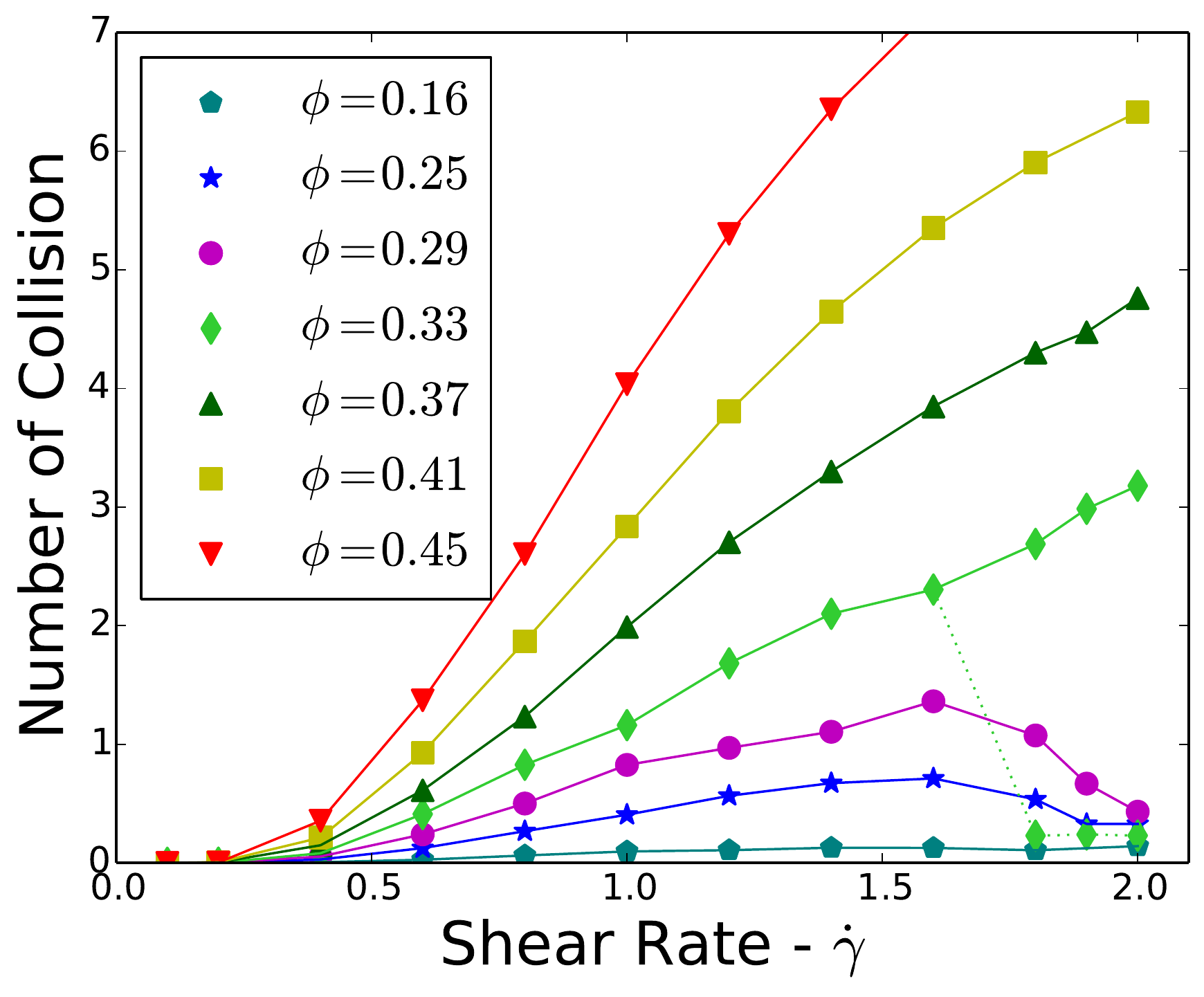}
\caption{\label{fig:colvis} Average number of nanoparticle collisions per
  frame as a function of shear rate for different volume
  fractions. The dashed line indicate the bifurcation for the
  intermediate volume fraction. The solid lines serve as a guide to eye.} 
\end{figure}

In this section we analyze the effect of nanoparticle-nanoparticle collisions on the 
viscosity. A collision between two nanoparticles is defined as an
interaction in which two atoms of the nanoparticles are closer than
$1\ \sigma$, which corresponds to the repulsive part of the interaction
potential. We depict in Fig. \ref{fig:colvis}(a) the number of collisions
per frame averaged over all the frames. 
At equilibrium and low shear
rates the repulsive part of the Lennard-Jones potential does not allow
particles to get too close, thus the number of collisions is
small. Moreover, if the volume fraction in nanoparticles is small, the
probability of finding two nanoparticles in the same neighborhood is
small and thus very few collisions occur. As the shear rate increases, the
increased stress results in more and more nanoparticle
collisions. At the same time we observe the collisions become more
violent with a larger repulsive force on average, thus resulting in a
larger energy dissipation and viscosity. 

However, we notice that for the volume fractions
exhibiting shear thinning  as the shear rate increases the number of collisions
decreases.  This observation is in agreement with the formation of the
sliding layers in systems exhibiting shear thinning. The layers permit the nanoparticles to avoid
colliding, and as a consequence the overall dissipation is reduced and
with it the shear viscosity. For the intermediate volume fraction
$\phi=0.33$ we observe the same behavior as the previous results, the number of collisions bifurcates into two
solutions, either a large number of collisions in the disordered state either a low number in the
ordered one. We must point out that we did not observe any collisions
involving more than two nanoparticles in our simulations, the only events are pairs of
nanoparticle colliding, as a consequence we can not attribute the increase in interactions to the formation of hydroclusters.

 \section{\label{sec5} Conclusion}

\begin{figure}[h!]
\includegraphics[scale=0.45]{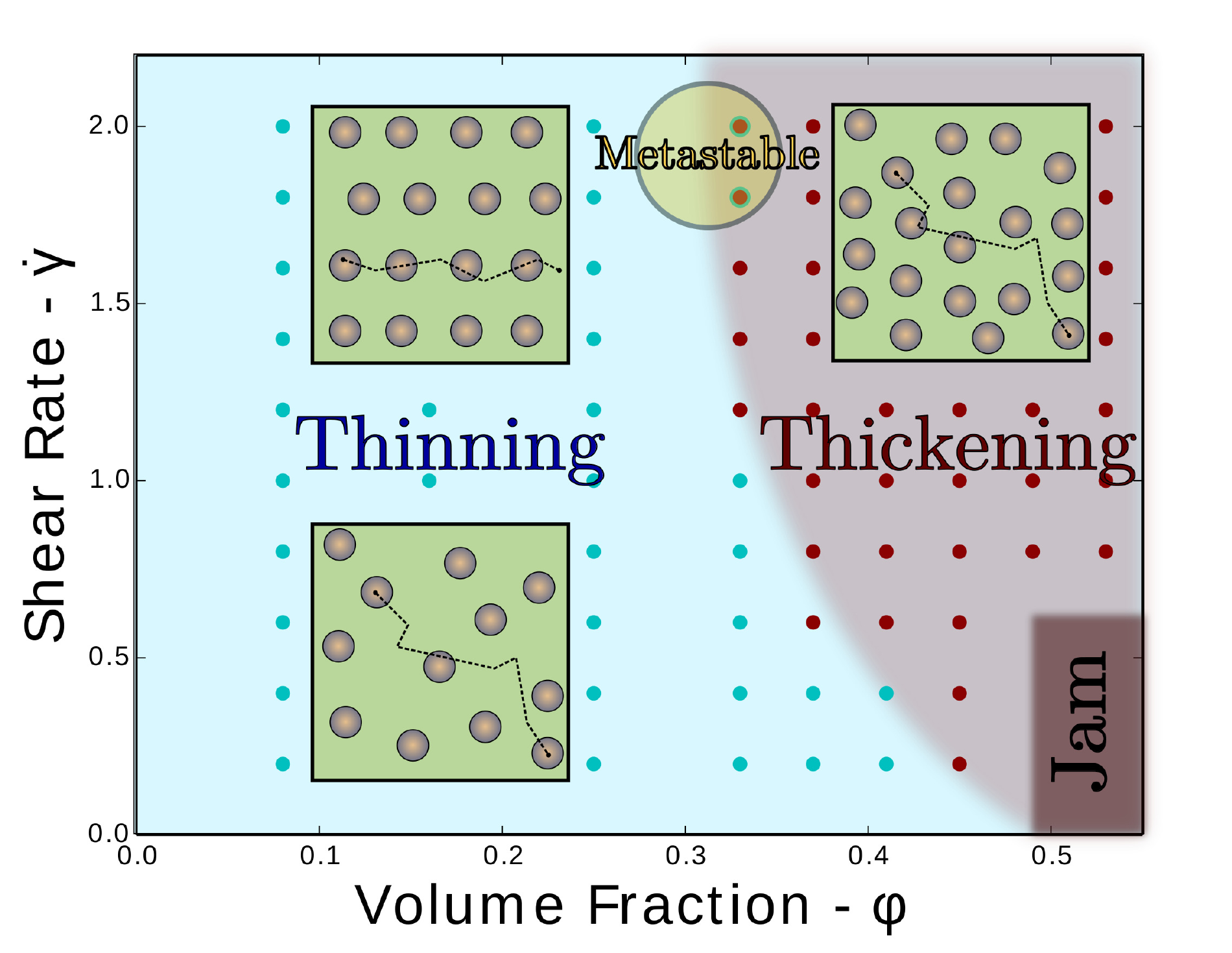}
\caption{\label{fig:phadia} Phase diagram of the flow
  characteristic as a function of shear rate and volume fraction. The circles represent simulation points.}
\end{figure}

We can summarize the results of the previous sections in a phase diagram as depicted in Fig. \ref{fig:phadia}.
At low volume fractions the nanoparticles are disordered and shear
thinning is observed, as the shear rate increases the nanoparticles
start to form layers in order to minimize collisions, and consequently
shear thinning is observed. As the volume fraction is further
increased we observe a metastable region where the nanoparticles go
through successions of ordered and disordered phases. The simulations
suggest that the duration of the ordered state increases with shear
rate however further research should be carried out. When the volume
fraction is increased further, there is not enough room for the
nanoparticles to form an ordered state, this in turn results in shear 
thickening. Finally, for very high volume fraction we observe a jammed
state at low shear rates, the fluid behaves as a solid, shear flow is
only possible after a sufficiently large shear rate is applied.  

Our simulations suggest that for a dispersion of spherical nanoparticles in a
simple fluid shear thickening is the result of the increased
collisions, and therefore energy dissipation. While for low shear rates, hydrodynamic and
Brownian forces are not enough to push the dispersed molecules into
the repulsive part of the interaction potential, the high shear rates force the nanoparticles
to collide. The commonly accepted mechanism of hydro-clustering
\cite{Pan,Guy,Mari,Heussinger,Seto,Denn,Fernandez2013}  is not the
leading mechanism for shear thickening in the dispersion of spherical
nanoparticles we consider in this study. We believe that one should
consider larger colloids--at scales of micrometers instead of
nanometers-- so that surface asperities and friction would contribute to
their formation. On the other hand, at high volume fractions the
leading mechanism for the transition from thinning to thickening
corresponds to the order-disorder transition previously suggested by
Hoffman \cite{Hoffman72,Hoffman98}. Our study suggests that there is
no single explanation for the shear thinning to thickening transition,
different mechanisms become dominant depending on the scale of the
colloids. However, the simple numerical model we propose permit to have a
microscopic understanding of the phenomenon.

\acknowledgements
This research is financially supported by the Istanbul Technical
University Scientific Research Fund (ITU-BAP) under Grant No. 38062.


\begin{thebibliography}{61}%
\makeatletter
\providecommand \@ifxundefined [1]{%
 \@ifx{#1\undefined}
}%
\providecommand \@ifnum [1]{%
 \ifnum #1\expandafter \@firstoftwo
 \else \expandafter \@secondoftwo
 \fi
}%
\providecommand \@ifx [1]{%
 \ifx #1\expandafter \@firstoftwo
 \else \expandafter \@secondoftwo
 \fi
}%
\providecommand \natexlab [1]{#1}%
\providecommand \enquote  [1]{``#1''}%
\providecommand \bibnamefont  [1]{#1}%
\providecommand \bibfnamefont [1]{#1}%
\providecommand \citenamefont [1]{#1}%
\providecommand \href@noop [0]{\@secondoftwo}%
\providecommand \href [0]{\begingroup \@sanitize@url \@href}%
\providecommand \@href[1]{\@@startlink{#1}\@@href}%
\providecommand \@@href[1]{\endgroup#1\@@endlink}%
\providecommand \@sanitize@url [0]{\catcode `\\12\catcode `\$12\catcode
  `\&12\catcode `\#12\catcode `\^12\catcode `\_12\catcode `\%12\relax}%
\providecommand \@@startlink[1]{}%
\providecommand \@@endlink[0]{}%
\providecommand \url  [0]{\begingroup\@sanitize@url \@url }%
\providecommand \@url [1]{\endgroup\@href {#1}{\urlprefix }}%
\providecommand \urlprefix  [0]{URL }%
\providecommand \Eprint [0]{\href }%
\providecommand \doibase [0]{http://dx.doi.org/}%
\providecommand \selectlanguage [0]{\@gobble}%
\providecommand \bibinfo  [0]{\@secondoftwo}%
\providecommand \bibfield  [0]{\@secondoftwo}%
\providecommand \translation [1]{[#1]}%
\providecommand \BibitemOpen [0]{}%
\providecommand \bibitemStop [0]{}%
\providecommand \bibitemNoStop [0]{.\EOS\space}%
\providecommand \EOS [0]{\spacefactor3000\relax}%
\providecommand \BibitemShut  [1]{\csname bibitem#1\endcsname}%
\let\auto@bib@innerbib\@empty
\bibitem [{\citenamefont {Katz}(2010)}]{JKatz}%
  \BibitemOpen
  \bibfield  {author} {\bibinfo {author} {\bibfnamefont {J.}~\bibnamefont
  {Katz}},\ }\href@noop {} {\emph {\bibinfo {title} {Introductory Fluid
  Mechanics}}}\ (\bibinfo  {publisher} {Cambridge University Press},\ \bibinfo
  {year} {2010})\BibitemShut {NoStop}%
\bibitem [{\citenamefont {Doi}\ and\ \citenamefont {Edwards}(1986)}]{Doi}%
  \BibitemOpen
  \bibfield  {author} {\bibinfo {author} {\bibfnamefont {M.}~\bibnamefont
  {Doi}}\ and\ \bibinfo {author} {\bibfnamefont {S.}~\bibnamefont {Edwards}},\
  }\href@noop {} {\emph {\bibinfo {title} {The Theory of Polymer Dynamics}}}\
  (\bibinfo  {publisher} {Clarendon Press, Oxford},\ \bibinfo {year}
  {1986})\BibitemShut {NoStop}%
\bibitem [{\citenamefont {Han}(2007)}]{Chang}%
  \BibitemOpen
  \bibfield  {author} {\bibinfo {author} {\bibfnamefont {C.~D.}\ \bibnamefont
  {Han}},\ }\href@noop {} {\emph {\bibinfo {title} {Rheology and Processing of
  Polymeric Materials,Volume 1: Polymer Rheology}}}\ (\bibinfo  {publisher}
  {Oxford University Press},\ \bibinfo {year} {2007})\BibitemShut {NoStop}%
\bibitem [{\citenamefont {Brown}\ and\ \citenamefont
  {Jaeger}(2014)}]{Jaeger14}%
  \BibitemOpen
  \bibfield  {author} {\bibinfo {author} {\bibfnamefont {E.}~\bibnamefont
  {Brown}}\ and\ \bibinfo {author} {\bibfnamefont {H.~M.}\ \bibnamefont
  {Jaeger}},\ }\href@noop {} {\bibfield  {journal} {\bibinfo  {journal} {Rep.
  Prog. Phys.}\ }\textbf {\bibinfo {volume} {77}},\ \bibinfo {pages} {046602}
  (\bibinfo {year} {2014})}\BibitemShut {NoStop}%
\bibitem [{\citenamefont {Bounoua}\ \emph {et~al.}(2016)\citenamefont
  {Bounoua}, \citenamefont {Lemaire}, \citenamefont {Férec}, \citenamefont
  {Ausias},\ and\ \citenamefont {Kuzhir}}]{bounoua16}%
  \BibitemOpen
  \bibfield  {author} {\bibinfo {author} {\bibfnamefont {S.}~\bibnamefont
  {Bounoua}}, \bibinfo {author} {\bibfnamefont {E.}~\bibnamefont {Lemaire}},
  \bibinfo {author} {\bibfnamefont {J.}~\bibnamefont {Férec}}, \bibinfo
  {author} {\bibfnamefont {G.}~\bibnamefont {Ausias}}, \ and\ \bibinfo {author}
  {\bibfnamefont {P.}~\bibnamefont {Kuzhir}},\ }\href@noop {} {\bibfield
  {journal} {\bibinfo  {journal} {J. Rheol.}\ }\textbf {\bibinfo {volume}
  {60}},\ \bibinfo {pages} {1279} (\bibinfo {year} {2016})}\BibitemShut
  {NoStop}%
\bibitem [{\citenamefont {Kapoor}, \citenamefont {Amandeep},\ and\
  \citenamefont {Patil}(2014)}]{Kapoor14}%
  \BibitemOpen
  \bibfield  {author} {\bibinfo {author} {\bibfnamefont {K.}~\bibnamefont
  {Kapoor}}, \bibinfo {author} {\bibnamefont {Amandeep}}, \ and\ \bibinfo
  {author} {\bibfnamefont {S.}~\bibnamefont {Patil}},\ }\href@noop {}
  {\bibfield  {journal} {\bibinfo  {journal} {Phys. Rev. E}\ }\textbf {\bibinfo
  {volume} {89}},\ \bibinfo {pages} {013004} (\bibinfo {year}
  {2014})}\BibitemShut {NoStop}%
\bibitem [{\citenamefont {Barnes}(1989)}]{Barnes89}%
  \BibitemOpen
  \bibfield  {author} {\bibinfo {author} {\bibfnamefont {H.~A.}\ \bibnamefont
  {Barnes}},\ }\href@noop {} {\bibfield  {journal} {\bibinfo  {journal} {J.
  Rheol.}\ }\textbf {\bibinfo {volume} {33}},\ \bibinfo {pages} {329} (\bibinfo
  {year} {1989})}\BibitemShut {NoStop}%
\bibitem [{\citenamefont {Crawford}\ \emph {et~al.}(2013)\citenamefont
  {Crawford}, \citenamefont {Popp}, \citenamefont {Johns}, \citenamefont
  {Caire}, \citenamefont {Peterson},\ and\ \citenamefont
  {Liberatore}}]{Crawford2013}%
  \BibitemOpen
  \bibfield  {author} {\bibinfo {author} {\bibfnamefont {N.~C.}\ \bibnamefont
  {Crawford}}, \bibinfo {author} {\bibfnamefont {L.~B.}\ \bibnamefont {Popp}},
  \bibinfo {author} {\bibfnamefont {K.~E.}\ \bibnamefont {Johns}}, \bibinfo
  {author} {\bibfnamefont {L.~M.}\ \bibnamefont {Caire}}, \bibinfo {author}
  {\bibfnamefont {B.~N.}\ \bibnamefont {Peterson}}, \ and\ \bibinfo {author}
  {\bibfnamefont {M.~W.}\ \bibnamefont {Liberatore}},\ }\href@noop {}
  {\bibfield  {journal} {\bibinfo  {journal} {J. Colloid Interface Sci.}\
  }\textbf {\bibinfo {volume} {396}},\ \bibinfo {pages} {83 } (\bibinfo {year}
  {2013})}\BibitemShut {NoStop}%
\bibitem [{\citenamefont {Hoffman}(1972)}]{Hoffman72}%
  \BibitemOpen
  \bibfield  {author} {\bibinfo {author} {\bibfnamefont {R.~L.}\ \bibnamefont
  {Hoffman}},\ }\href@noop {} {\bibfield  {journal} {\bibinfo  {journal} {Trans. Soc.
  Rheol.}\ }\textbf {\bibinfo {volume} {16}},\ \bibinfo {pages} {155} (\bibinfo
  {year} {1972})}\BibitemShut {NoStop}%
\bibitem [{\citenamefont {Hoffman}(1974)}]{Hoffman74}%
  \BibitemOpen
  \bibfield  {author} {\bibinfo {author} {\bibfnamefont {R.}~\bibnamefont
  {Hoffman}},\ }\href@noop {} {\bibfield  {journal} {\bibinfo  {journal} {J.
  Colloid Interface Sci.}\ }\textbf {\bibinfo {volume} {46}},\ \bibinfo {pages}
  {491 } (\bibinfo {year} {1974})}\BibitemShut {NoStop}%
\bibitem [{\citenamefont {Hoffman}(1991)}]{hoffman1991}%
  \BibitemOpen
  \bibfield  {author} {\bibinfo {author} {\bibfnamefont {R.~L.}\ \bibnamefont
  {Hoffman}},\ }\href@noop {} {\bibfield  {journal} {\bibinfo  {journal} {MRS
  Bull.}\ }\textbf {\bibinfo {volume} {16}},\ \bibinfo {pages} {32–37}
  (\bibinfo {year} {1991})}\BibitemShut {NoStop}%
\bibitem [{\citenamefont {Bender}\ and\ \citenamefont
  {Wagner}(1996)}]{wagner96}%
  \BibitemOpen
  \bibfield  {author} {\bibinfo {author} {\bibfnamefont {J.}~\bibnamefont
  {Bender}}\ and\ \bibinfo {author} {\bibfnamefont {N.~J.}\ \bibnamefont
  {Wagner}},\ }\href@noop {} {\bibfield  {journal} {\bibinfo  {journal} {J.
  Rheol.}\ }\textbf {\bibinfo {volume} {40}},\ \bibinfo {pages} {899} (\bibinfo
  {year} {1996})}\BibitemShut {NoStop}%
\bibitem [{\citenamefont {Madraki}\ \emph {et~al.}(2017)\citenamefont
  {Madraki}, \citenamefont {Hormozi}, \citenamefont {Ovarlez}, \citenamefont
  {Guazzelli},\ and\ \citenamefont {Pouliquen}}]{Madraki2017}%
  \BibitemOpen
  \bibfield  {author} {\bibinfo {author} {\bibfnamefont {Y.}~\bibnamefont
  {Madraki}}, \bibinfo {author} {\bibfnamefont {S.}~\bibnamefont {Hormozi}},
  \bibinfo {author} {\bibfnamefont {G.}~\bibnamefont {Ovarlez}}, \bibinfo
  {author} {\bibfnamefont {E.}~\bibnamefont {Guazzelli}}, \ and\ \bibinfo
  {author} {\bibfnamefont {O.}~\bibnamefont {Pouliquen}},\ }\href@noop {}
  {\bibfield  {journal} {\bibinfo  {journal} {Phys. Rev. Fluids}\ }\textbf
  {\bibinfo {volume} {2}},\ \bibinfo {pages} {033301} (\bibinfo {year}
  {2017})}\BibitemShut {NoStop}%
\bibitem [{\citenamefont {Maybury}, \citenamefont {Ho},\ and\ \citenamefont
  {Binhowimal}(2017)}]{Cement}%
  \BibitemOpen
  \bibfield  {author} {\bibinfo {author} {\bibfnamefont {J.}~\bibnamefont
  {Maybury}}, \bibinfo {author} {\bibfnamefont {J.}~\bibnamefont {Ho}}, \ and\
  \bibinfo {author} {\bibfnamefont {S.}~\bibnamefont {Binhowimal}},\
  }\href@noop {} {\bibfield  {journal} {\bibinfo  {journal} {Constr Build
  Mater}\ }\textbf {\bibinfo {volume} {142}},\ \bibinfo {pages} {268 }
  (\bibinfo {year} {2017})}\BibitemShut {NoStop}%
\bibitem [{\citenamefont {Khandavalli}\ and\ \citenamefont
  {Rothstein}(2016)}]{Dye}%
  \BibitemOpen
  \bibfield  {author} {\bibinfo {author} {\bibfnamefont {S.}~\bibnamefont
  {Khandavalli}}\ and\ \bibinfo {author} {\bibfnamefont {J.~P.}\ \bibnamefont
  {Rothstein}},\ }\href@noop {} {\bibfield  {journal} {\bibinfo  {journal}
  {AIChE J.}\ }\textbf {\bibinfo {volume} {62}},\ \bibinfo {pages} {4536}
  (\bibinfo {year} {2016})}\BibitemShut {NoStop}%
\bibitem [{\citenamefont {Lee}, \citenamefont {Wetzel},\ and\ \citenamefont
  {Wagner}(2003)}]{Lee2003}%
  \BibitemOpen
  \bibfield  {author} {\bibinfo {author} {\bibfnamefont {Y.~S.}\ \bibnamefont
  {Lee}}, \bibinfo {author} {\bibfnamefont {E.~D.}\ \bibnamefont {Wetzel}}, \
  and\ \bibinfo {author} {\bibfnamefont {N.~J.}\ \bibnamefont {Wagner}},\
  }\href@noop {} {\bibfield  {journal} {\bibinfo  {journal} {J. Mater. Sci.}\
  }\textbf {\bibinfo {volume} {38}},\ \bibinfo {pages} {2825} (\bibinfo {year}
  {2003})}\BibitemShut {NoStop}%
\bibitem [{\citenamefont {Gong}\ \emph {et~al.}(2014)\citenamefont {Gong},
  \citenamefont {Xu}, \citenamefont {Zhu}, \citenamefont {Xuan}, \citenamefont
  {Jiang},\ and\ \citenamefont {Jiang}}]{Gong2014}%
  \BibitemOpen
  \bibfield  {author} {\bibinfo {author} {\bibfnamefont {X.}~\bibnamefont
  {Gong}}, \bibinfo {author} {\bibfnamefont {Y.}~\bibnamefont {Xu}}, \bibinfo
  {author} {\bibfnamefont {W.}~\bibnamefont {Zhu}}, \bibinfo {author}
  {\bibfnamefont {S.}~\bibnamefont {Xuan}}, \bibinfo {author} {\bibfnamefont
  {W.}~\bibnamefont {Jiang}}, \ and\ \bibinfo {author} {\bibfnamefont
  {W.}~\bibnamefont {Jiang}},\ }\href@noop {} {\bibfield  {journal} {\bibinfo
  {journal} {J. Compos. Mater.}\ }\textbf {\bibinfo {volume} {48}},\ \bibinfo
  {pages} {641} (\bibinfo {year} {2014})}\BibitemShut {NoStop}%
\bibitem [{\citenamefont {Li}\ \emph {et~al.}(2014)\citenamefont {Li},
  \citenamefont {Wang}, \citenamefont {Ding}, \citenamefont {Wu},\ and\
  \citenamefont {Fu}}]{Li2014}%
  \BibitemOpen
  \bibfield  {author} {\bibinfo {author} {\bibfnamefont {S.}~\bibnamefont
  {Li}}, \bibinfo {author} {\bibfnamefont {Y.}~\bibnamefont {Wang}}, \bibinfo
  {author} {\bibfnamefont {J.}~\bibnamefont {Ding}}, \bibinfo {author}
  {\bibfnamefont {H.}~\bibnamefont {Wu}}, \ and\ \bibinfo {author}
  {\bibfnamefont {Y.}~\bibnamefont {Fu}},\ }\href@noop {} {\bibfield  {journal}
  {\bibinfo  {journal} {Text. Res. J.}\ }\textbf {\bibinfo {volume} {84}},\
  \bibinfo {pages} {897} (\bibinfo {year} {2014})}\BibitemShut {NoStop}%
\bibitem [{\citenamefont {Jiang}\ \emph {et~al.}(2014)\citenamefont {Jiang},
  \citenamefont {Ye}, \citenamefont {He}, \citenamefont {Gong}, \citenamefont
  {Feng}, \citenamefont {Lu},\ and\ \citenamefont {Xuan}}]{Jiang2014}%
  \BibitemOpen
  \bibfield  {author} {\bibinfo {author} {\bibfnamefont {W.}~\bibnamefont
  {Jiang}}, \bibinfo {author} {\bibfnamefont {F.}~\bibnamefont {Ye}}, \bibinfo
  {author} {\bibfnamefont {Q.}~\bibnamefont {He}}, \bibinfo {author}
  {\bibfnamefont {X.}~\bibnamefont {Gong}}, \bibinfo {author} {\bibfnamefont
  {J.}~\bibnamefont {Feng}}, \bibinfo {author} {\bibfnamefont {L.}~\bibnamefont
  {Lu}}, \ and\ \bibinfo {author} {\bibfnamefont {S.}~\bibnamefont {Xuan}},\
  }\href@noop {} {\bibfield  {journal} {\bibinfo  {journal} {J. Colloid
  Interface Sci.}\ }\textbf {\bibinfo {volume} {413}},\ \bibinfo {pages} {8 }
  (\bibinfo {year} {2014})}\BibitemShut {NoStop}%
\bibitem [{\citenamefont {Chow}\ and\ \citenamefont
  {Zukoski}(1995)}]{Chow1995}%
  \BibitemOpen
  \bibfield  {author} {\bibinfo {author} {\bibfnamefont {M.~K.}\ \bibnamefont
  {Chow}}\ and\ \bibinfo {author} {\bibfnamefont {C.~F.}\ \bibnamefont
  {Zukoski}},\ }\href@noop {} {\bibfield  {journal} {\bibinfo  {journal} {J.
  Rheol.}\ }\textbf {\bibinfo {volume} {39}},\ \bibinfo {pages} {33} (\bibinfo
  {year} {1995})}\BibitemShut {NoStop}%
\bibitem [{\citenamefont {Fall}\ \emph {et~al.}(2015)\citenamefont {Fall},
  \citenamefont {Bertrand}, \citenamefont {Hautemayou}, \citenamefont
  {Mezi\`ere}, \citenamefont {Moucheront}, \citenamefont {Lema\^{\i}tre},\ and\
  \citenamefont {Ovarlez}}]{Fall2015}%
  \BibitemOpen
  \bibfield  {author} {\bibinfo {author} {\bibfnamefont {A.}~\bibnamefont
  {Fall}}, \bibinfo {author} {\bibfnamefont {F.}~\bibnamefont {Bertrand}},
  \bibinfo {author} {\bibfnamefont {D.}~\bibnamefont {Hautemayou}}, \bibinfo
  {author} {\bibfnamefont {C.}~\bibnamefont {Mezi\`ere}}, \bibinfo {author}
  {\bibfnamefont {P.}~\bibnamefont {Moucheront}}, \bibinfo {author}
  {\bibfnamefont {A.}~\bibnamefont {Lema\^{\i}tre}}, \ and\ \bibinfo {author}
  {\bibfnamefont {G.}~\bibnamefont {Ovarlez}},\ }\href@noop {} {\bibfield
  {journal} {\bibinfo  {journal} {Phys. Rev. Lett.}\ }\textbf {\bibinfo
  {volume} {114}},\ \bibinfo {pages} {098301} (\bibinfo {year}
  {2015})}\BibitemShut {NoStop}%
\bibitem [{\citenamefont {Cwalina}, \citenamefont {Harrison},\ and\
  \citenamefont {Wagner}(2016)}]{Cwalina2016}%
  \BibitemOpen
  \bibfield  {author} {\bibinfo {author} {\bibfnamefont {D.~C.}\ \bibnamefont
  {Cwalina}}, \bibinfo {author} {\bibfnamefont {J.~K.}\ \bibnamefont
  {Harrison}}, \ and\ \bibinfo {author} {\bibfnamefont {J.~N.}\ \bibnamefont
  {Wagner}},\ }\href@noop {} {\bibfield  {journal} {\bibinfo  {journal} {Soft
  Matter}\ }\textbf {\bibinfo {volume} {12}},\ \bibinfo {pages} {4654}
  (\bibinfo {year} {2016})}\BibitemShut {NoStop}%
\bibitem [{\citenamefont {Foss}\ and\ \citenamefont
  {Brady}(2000)}]{foss_brady_2000}%
  \BibitemOpen
  \bibfield  {author} {\bibinfo {author} {\bibfnamefont {D.~R.}\ \bibnamefont
  {Foss}}\ and\ \bibinfo {author} {\bibfnamefont {J.~F.}\ \bibnamefont
  {Brady}},\ }\href@noop {} {\bibfield  {journal} {\bibinfo  {journal} {J.
  Fluid Mech.}\ }\textbf {\bibinfo {volume} {407}},\ \bibinfo {pages}
  {167–200} (\bibinfo {year} {2000})}\BibitemShut {NoStop}%
\bibitem [{\citenamefont {Seto}\ \emph
  {et~al.}(2013{\natexlab{a}})\citenamefont {Seto}, \citenamefont {Mari},
  \citenamefont {Morris},\ and\ \citenamefont {Denn}}]{Seto2013}%
  \BibitemOpen
  \bibfield  {author} {\bibinfo {author} {\bibfnamefont {R.}~\bibnamefont
  {Seto}}, \bibinfo {author} {\bibfnamefont {R.}~\bibnamefont {Mari}}, \bibinfo
  {author} {\bibfnamefont {J.~F.}\ \bibnamefont {Morris}}, \ and\ \bibinfo
  {author} {\bibfnamefont {M.~M.}\ \bibnamefont {Denn}},\ }\href@noop {}
  {\bibfield  {journal} {\bibinfo  {journal} {Phys. Rev. Lett.}\ }\textbf
  {\bibinfo {volume} {111}},\ \bibinfo {pages} {218301} (\bibinfo {year}
  {2013}{\natexlab{a}})}\BibitemShut {NoStop}%
\bibitem [{\citenamefont {Mari}\ \emph {et~al.}(2015)\citenamefont {Mari},
  \citenamefont {Seto}, \citenamefont {Morris},\ and\ \citenamefont
  {Denn}}]{Mari2015}%
  \BibitemOpen
  \bibfield  {author} {\bibinfo {author} {\bibfnamefont {R.}~\bibnamefont
  {Mari}}, \bibinfo {author} {\bibfnamefont {R.}~\bibnamefont {Seto}}, \bibinfo
  {author} {\bibfnamefont {J.~F.}\ \bibnamefont {Morris}}, \ and\ \bibinfo
  {author} {\bibfnamefont {M.~M.}\ \bibnamefont {Denn}},\ }\href@noop {}
  {\bibfield  {journal} {\bibinfo  {journal} {Proc. Natl. Acad. Sci. U.S.A.}\
  }\textbf {\bibinfo {volume} {112}},\ \bibinfo {pages} {15326} (\bibinfo
  {year} {2015})}\BibitemShut {NoStop}%
\bibitem [{\citenamefont {Pednekar}, \citenamefont {Chun},\ and\ \citenamefont
  {Morris}(2017)}]{Pednekar2017}%
  \BibitemOpen
  \bibfield  {author} {\bibinfo {author} {\bibfnamefont {S.}~\bibnamefont
  {Pednekar}}, \bibinfo {author} {\bibfnamefont {J.}~\bibnamefont {Chun}}, \
  and\ \bibinfo {author} {\bibfnamefont {J.~F.}\ \bibnamefont {Morris}},\
  }\href@noop {} {\bibfield  {journal} {\bibinfo  {journal} {Soft Matter}\
  }\textbf {\bibinfo {volume} {13}},\ \bibinfo {pages} {1773} (\bibinfo {year}
  {2017})}\BibitemShut {NoStop}%
\bibitem [{\citenamefont {Maranzano}\ and\ \citenamefont
  {Wagner}(2001)}]{Wagner2001}%
  \BibitemOpen
  \bibfield  {author} {\bibinfo {author} {\bibfnamefont {B.~J.}\ \bibnamefont
  {Maranzano}}\ and\ \bibinfo {author} {\bibfnamefont {N.~J.}\ \bibnamefont
  {Wagner}},\ }\href@noop {} {\bibfield  {journal} {\bibinfo  {journal} {J.
  Chem. Phys.}\ }\textbf {\bibinfo {volume} {114}},\ \bibinfo {pages} {10514}
  (\bibinfo {year} {2001})}\BibitemShut {NoStop}%
\bibitem [{\citenamefont {Li}\ \emph {et~al.}(2017)\citenamefont {Li},
  \citenamefont {Wang}, \citenamefont {Zhao}, \citenamefont {Cai},
  \citenamefont {Wang},\ and\ \citenamefont {Wang}}]{Wang2017}%
  \BibitemOpen
  \bibfield  {author} {\bibinfo {author} {\bibfnamefont {S.}~\bibnamefont
  {Li}}, \bibinfo {author} {\bibfnamefont {J.}~\bibnamefont {Wang}}, \bibinfo
  {author} {\bibfnamefont {S.}~\bibnamefont {Zhao}}, \bibinfo {author}
  {\bibfnamefont {W.}~\bibnamefont {Cai}}, \bibinfo {author} {\bibfnamefont
  {Z.}~\bibnamefont {Wang}}, \ and\ \bibinfo {author} {\bibfnamefont
  {S.}~\bibnamefont {Wang}},\ }\href@noop {} {\bibfield  {journal} {\bibinfo
  {journal} {J Mater Sci Technol}\ }\textbf {\bibinfo {volume} {33}},\ \bibinfo
  {pages} {261 } (\bibinfo {year} {2017})}\BibitemShut {NoStop}%
\bibitem [{\citenamefont {Wolthers}\ \emph {et~al.}(1996)\citenamefont
  {Wolthers}, \citenamefont {van~den Ende}, \citenamefont {Duits},\ and\
  \citenamefont {Mellema}}]{Wolthers}%
  \BibitemOpen
  \bibfield  {author} {\bibinfo {author} {\bibfnamefont {W.}~\bibnamefont
  {Wolthers}}, \bibinfo {author} {\bibfnamefont {D.}~\bibnamefont {van~den
  Ende}}, \bibinfo {author} {\bibfnamefont {M.~H.~G.}\ \bibnamefont {Duits}}, \
  and\ \bibinfo {author} {\bibfnamefont {J.}~\bibnamefont {Mellema}},\
  }\href@noop {} {\bibfield  {journal} {\bibinfo  {journal} {J. Rheol.}\
  }\textbf {\bibinfo {volume} {40}},\ \bibinfo {pages} {55} (\bibinfo {year}
  {1996})}\BibitemShut {NoStop}%
\bibitem [{\citenamefont {Peters}, \citenamefont {Majumdar},\ and\
  \citenamefont {Jaeger}(2016)}]{Peters2016}%
  \BibitemOpen
  \bibfield  {author} {\bibinfo {author} {\bibfnamefont {I.~R.}\ \bibnamefont
  {Peters}}, \bibinfo {author} {\bibfnamefont {S.}~\bibnamefont {Majumdar}}, \
  and\ \bibinfo {author} {\bibfnamefont {H.~M.}\ \bibnamefont {Jaeger}},\
  }\href@noop {} {\bibfield  {journal} {\bibinfo  {journal} {Nature}\ }\textbf
  {\bibinfo {volume} {532}},\ \bibinfo {pages} {214} (\bibinfo {year}
  {2016})}\BibitemShut {NoStop}%
\bibitem [{\citenamefont {Fall}\ \emph {et~al.}(2008)\citenamefont {Fall},
  \citenamefont {Huang}, \citenamefont {Bertrand}, \citenamefont {Ovarlez},\
  and\ \citenamefont {Bonn}}]{Fall2008}%
  \BibitemOpen
  \bibfield  {author} {\bibinfo {author} {\bibfnamefont {A.}~\bibnamefont
  {Fall}}, \bibinfo {author} {\bibfnamefont {N.}~\bibnamefont {Huang}},
  \bibinfo {author} {\bibfnamefont {F.}~\bibnamefont {Bertrand}}, \bibinfo
  {author} {\bibfnamefont {G.}~\bibnamefont {Ovarlez}}, \ and\ \bibinfo
  {author} {\bibfnamefont {D.}~\bibnamefont {Bonn}},\ }\href@noop {} {\bibfield
   {journal} {\bibinfo  {journal} {Phys. Rev. Lett.}\ }\textbf {\bibinfo
  {volume} {100}},\ \bibinfo {pages} {018301} (\bibinfo {year}
  {2008})}\BibitemShut {NoStop}%
\bibitem [{\citenamefont {Brown}\ \emph {et~al.}(2010)\citenamefont {Brown},
  \citenamefont {Forman}, \citenamefont {Orellana}, \citenamefont {Zhang},
  \citenamefont {Maynor}, \citenamefont {Betts}, \citenamefont {DeSimone},\
  and\ \citenamefont {Jaeger}}]{Jaeger10}%
  \BibitemOpen
  \bibfield  {author} {\bibinfo {author} {\bibfnamefont {E.}~\bibnamefont
  {Brown}}, \bibinfo {author} {\bibfnamefont {N.~A.}\ \bibnamefont {Forman}},
  \bibinfo {author} {\bibfnamefont {C.~S.}\ \bibnamefont {Orellana}}, \bibinfo
  {author} {\bibfnamefont {H.}~\bibnamefont {Zhang}}, \bibinfo {author}
  {\bibfnamefont {B.~W.}\ \bibnamefont {Maynor}}, \bibinfo {author}
  {\bibfnamefont {D.~E.}\ \bibnamefont {Betts}}, \bibinfo {author}
  {\bibfnamefont {J.~M.}\ \bibnamefont {DeSimone}}, \ and\ \bibinfo {author}
  {\bibfnamefont {H.~M.}\ \bibnamefont {Jaeger}},\ }\href@noop {} {\bibfield
  {journal} {\bibinfo  {journal} {Nat. Mater.}\ }\textbf {\bibinfo {volume}
  {9}},\ \bibinfo {pages} {220} (\bibinfo {year} {2010})}\BibitemShut {NoStop}%
\bibitem [{\citenamefont {Hoffman}(1998)}]{Hoffman98}%
  \BibitemOpen
  \bibfield  {author} {\bibinfo {author} {\bibfnamefont {R.~L.}\ \bibnamefont
  {Hoffman}},\ }\href@noop {} {\bibfield  {journal} {\bibinfo  {journal} {J.
  Rheol.}\ }\textbf {\bibinfo {volume} {42}},\ \bibinfo {pages} {111} (\bibinfo
  {year} {1998})}\BibitemShut {NoStop}%
\bibitem [{\citenamefont {Ackerson}\ and\ \citenamefont
  {Pusey}(1988)}]{Ackerson1988}%
  \BibitemOpen
  \bibfield  {author} {\bibinfo {author} {\bibfnamefont {B.~J.}\ \bibnamefont
  {Ackerson}}\ and\ \bibinfo {author} {\bibfnamefont {P.~N.}\ \bibnamefont
  {Pusey}},\ }\href@noop {} {\bibfield  {journal} {\bibinfo  {journal} {Phys.
  Rev. Lett.}\ }\textbf {\bibinfo {volume} {61}},\ \bibinfo {pages} {1033}
  (\bibinfo {year} {1988})}\BibitemShut {NoStop}%
\bibitem [{\citenamefont {Yan}\ \emph {et~al.}(1994)\citenamefont {Yan},
  \citenamefont {Dhont}, \citenamefont {Smits},\ and\ \citenamefont
  {Lekkerkerker}}]{Yan1994}%
  \BibitemOpen
  \bibfield  {author} {\bibinfo {author} {\bibfnamefont {Y.~D.}\ \bibnamefont
  {Yan}}, \bibinfo {author} {\bibfnamefont {J.~K.~G.}\ \bibnamefont {Dhont}},
  \bibinfo {author} {\bibfnamefont {C.}~\bibnamefont {Smits}}, \ and\ \bibinfo
  {author} {\bibfnamefont {H.~N.~W.}\ \bibnamefont {Lekkerkerker}},\
  }\href@noop {} {\bibfield  {journal} {\bibinfo  {journal} {Physica A}\
  }\textbf {\bibinfo {volume} {202}},\ \bibinfo {pages} {68} (\bibinfo {year}
  {1994})}\BibitemShut {NoStop}%
\bibitem [{\citenamefont {Brady}\ and\ \citenamefont
  {Bossis}(1985)}]{brady_bossis1985}%
  \BibitemOpen
  \bibfield  {author} {\bibinfo {author} {\bibfnamefont {J.~F.}\ \bibnamefont
  {Brady}}\ and\ \bibinfo {author} {\bibfnamefont {G.}~\bibnamefont {Bossis}},\
  }\href@noop {} {\bibfield  {journal} {\bibinfo  {journal} {J. Fluid Mech.}\
  }\textbf {\bibinfo {volume} {155}},\ \bibinfo {pages} {105–129} (\bibinfo
  {year} {1985})}\BibitemShut {NoStop}%
\bibitem [{\citenamefont {Maranzano}\ and\ \citenamefont
  {Wagner}(2002)}]{maranzano2002}%
  \BibitemOpen
  \bibfield  {author} {\bibinfo {author} {\bibfnamefont {B.~J.}\ \bibnamefont
  {Maranzano}}\ and\ \bibinfo {author} {\bibfnamefont {N.~J.}\ \bibnamefont
  {Wagner}},\ }\href@noop {} {\bibfield  {journal} {\bibinfo  {journal} {J.
  Chem. Phys.}\ }\textbf {\bibinfo {volume} {117}},\ \bibinfo {pages} {10291}
  (\bibinfo {year} {2002})}\BibitemShut {NoStop}%
\bibitem [{\citenamefont {Kalman}\ and\ \citenamefont
  {Wagner}(2009)}]{Kalman2009}%
  \BibitemOpen
  \bibfield  {author} {\bibinfo {author} {\bibfnamefont {D.~P.}\ \bibnamefont
  {Kalman}}\ and\ \bibinfo {author} {\bibfnamefont {N.~J.}\ \bibnamefont
  {Wagner}},\ }\href@noop {} {\bibfield  {journal} {\bibinfo  {journal} {Rheol
  Acta}\ }\textbf {\bibinfo {volume} {48}},\ \bibinfo {pages} {897} (\bibinfo
  {year} {2009})}\BibitemShut {NoStop}%
\bibitem [{\citenamefont {Cheng}\ \emph {et~al.}(2011)\citenamefont {Cheng},
  \citenamefont {McCoy}, \citenamefont {Israelachvili},\ and\ \citenamefont
  {Cohen}}]{Cheng2011}%
  \BibitemOpen
  \bibfield  {author} {\bibinfo {author} {\bibfnamefont {X.}~\bibnamefont
  {Cheng}}, \bibinfo {author} {\bibfnamefont {J.~H.}\ \bibnamefont {McCoy}},
  \bibinfo {author} {\bibfnamefont {J.~N.}\ \bibnamefont {Israelachvili}}, \
  and\ \bibinfo {author} {\bibfnamefont {I.}~\bibnamefont {Cohen}},\
  }\href@noop {} {\bibfield  {journal} {\bibinfo  {journal} {Science}\ }\textbf
  {\bibinfo {volume} {333}},\ \bibinfo {pages} {1276} (\bibinfo {year}
  {2011})}\BibitemShut {NoStop}%
\bibitem [{\citenamefont {Dullweber}, \citenamefont {Leimkuhler},\ and\
  \citenamefont {McLachlan}(1997)}]{Dullweber}%
  \BibitemOpen
  \bibfield  {author} {\bibinfo {author} {\bibfnamefont {A.}~\bibnamefont
  {Dullweber}}, \bibinfo {author} {\bibfnamefont {B.}~\bibnamefont
  {Leimkuhler}}, \ and\ \bibinfo {author} {\bibfnamefont {R.}~\bibnamefont
  {McLachlan}},\ }\href@noop {} {\bibfield  {journal} {\bibinfo  {journal} {J.
  Chem. Phys.}\ }\textbf {\bibinfo {volume} {107}},\ \bibinfo {pages} {5840}
  (\bibinfo {year} {1997})}\BibitemShut {NoStop}%
\bibitem [{\citenamefont {Rapaport}(2004)}]{Rapaport}%
  \BibitemOpen
  \bibfield  {author} {\bibinfo {author} {\bibfnamefont {D.~C.}\ \bibnamefont
  {Rapaport}},\ }\href@noop {} {\emph {\bibinfo {title} {The Art of Molecular
  Dynamics Simulaton}}}\ (\bibinfo  {publisher} {Cambridge University Press},\
  \bibinfo {year} {2004})\BibitemShut {NoStop}%
\bibitem [{\citenamefont {Ryckaert}, \citenamefont {Ciccotti},\ and\
  \citenamefont {Berendsen}(1977)}]{SHAKE}%
  \BibitemOpen
  \bibfield  {author} {\bibinfo {author} {\bibfnamefont {J.-P.}\ \bibnamefont
  {Ryckaert}}, \bibinfo {author} {\bibfnamefont {G.}~\bibnamefont {Ciccotti}},
  \ and\ \bibinfo {author} {\bibfnamefont {H.~J.}\ \bibnamefont {Berendsen}},\
  }\href@noop {} {\bibfield  {journal} {\bibinfo  {journal} {J. Comput. Phys.}\
  }\textbf {\bibinfo {volume} {23}},\ \bibinfo {pages} {327 } (\bibinfo {year}
  {1977})}\BibitemShut {NoStop}%
\bibitem [{\citenamefont {Andersen}(1983)}]{RATTLE}%
  \BibitemOpen
  \bibfield  {author} {\bibinfo {author} {\bibfnamefont {H.~C.}\ \bibnamefont
  {Andersen}},\ }\href@noop {} {\bibfield  {journal} {\bibinfo  {journal} {J.
  Comput. Phys.}\ }\textbf {\bibinfo {volume} {52}},\ \bibinfo {pages} {24 }
  (\bibinfo {year} {1983})}\BibitemShut {NoStop}%
\bibitem [{\citenamefont {Sunarso}, \citenamefont {Tsuji},\ and\ \citenamefont
  {Chono}(2011)}]{Sunarso2011}%
  \BibitemOpen
  \bibfield  {author} {\bibinfo {author} {\bibfnamefont {A.}~\bibnamefont
  {Sunarso}}, \bibinfo {author} {\bibfnamefont {T.}~\bibnamefont {Tsuji}}, \
  and\ \bibinfo {author} {\bibfnamefont {S.}~\bibnamefont {Chono}},\
  }\href@noop {} {\bibfield  {journal} {\bibinfo  {journal} {J. Appl. Phys.}\
  }\textbf {\bibinfo {volume} {110}},\ \bibinfo {pages} {044911} (\bibinfo
  {year} {2011})}\BibitemShut {NoStop}%
\bibitem [{\citenamefont {Akimov}\ and\ \citenamefont
  {Kolomeisky}(2011)}]{Akimov2011}%
  \BibitemOpen
  \bibfield  {author} {\bibinfo {author} {\bibfnamefont {A.}~\bibnamefont
  {Akimov}}\ and\ \bibinfo {author} {\bibfnamefont {A.~B.}\ \bibnamefont
  {Kolomeisky}},\ }\href@noop {} {\bibfield  {journal} {\bibinfo  {journal} {J.
  Phys. Chem. C}\ }\textbf {\bibinfo {volume} {115}},\ \bibinfo {pages} {125}
  (\bibinfo {year} {2011})}\BibitemShut {NoStop}%
\bibitem [{\citenamefont {Orsi}, \citenamefont {Michel},\ and\ \citenamefont
  {Essex}(2010)}]{Orsi2010}%
  \BibitemOpen
  \bibfield  {author} {\bibinfo {author} {\bibfnamefont {M.}~\bibnamefont
  {Orsi}}, \bibinfo {author} {\bibfnamefont {J.}~\bibnamefont {Michel}}, \ and\
  \bibinfo {author} {\bibfnamefont {J.~W.}\ \bibnamefont {Essex}},\ }\href@noop
  {} {\bibfield  {journal} {\bibinfo  {journal} {J. Phys. Condens. Matter}\
  }\textbf {\bibinfo {volume} {22}},\ \bibinfo {pages} {155106} (\bibinfo
  {year} {2010})}\BibitemShut {NoStop}%
\bibitem [{\citenamefont {Lees}\ and\ \citenamefont
  {Edwards}(1972)}]{Lees_Edwards}%
  \BibitemOpen
  \bibfield  {author} {\bibinfo {author} {\bibfnamefont {A.~W.}\ \bibnamefont
  {Lees}}\ and\ \bibinfo {author} {\bibfnamefont {S.~F.}\ \bibnamefont
  {Edwards}},\ }\href@noop {} {\bibfield  {journal} {\bibinfo  {journal} {J
  Phys C Solid State}\ }\textbf {\bibinfo {volume} {5}},\ \bibinfo {pages}
  {1921} (\bibinfo {year} {1972})}\BibitemShut {NoStop}%
\bibitem [{\citenamefont {Edberg}, \citenamefont {Morriss},\ and\ \citenamefont
  {Evans}(1987)}]{Evans86}%
  \BibitemOpen
  \bibfield  {author} {\bibinfo {author} {\bibfnamefont {R.}~\bibnamefont
  {Edberg}}, \bibinfo {author} {\bibfnamefont {G.~P.}\ \bibnamefont {Morriss}},
  \ and\ \bibinfo {author} {\bibfnamefont {D.~J.}\ \bibnamefont {Evans}},\
  }\href@noop {} {\bibfield  {journal} {\bibinfo  {journal} {J. Chem. Phys.}\
  }\textbf {\bibinfo {volume} {86}},\ \bibinfo {pages} {4555} (\bibinfo {year}
  {1987})}\BibitemShut {NoStop}%
\bibitem [{\citenamefont {Travis}, \citenamefont {Daivis},\ and\ \citenamefont
  {Evans}(1995)}]{Evans95}%
  \BibitemOpen
  \bibfield  {author} {\bibinfo {author} {\bibfnamefont {K.~P.}\ \bibnamefont
  {Travis}}, \bibinfo {author} {\bibfnamefont {P.~J.}\ \bibnamefont {Daivis}},
  \ and\ \bibinfo {author} {\bibfnamefont {D.~J.}\ \bibnamefont {Evans}},\
  }\href@noop {} {\bibfield  {journal} {\bibinfo  {journal} {J. Chem. Phys.}\
  }\textbf {\bibinfo {volume} {103}},\ \bibinfo {pages} {1109} (\bibinfo {year}
  {1995})}\BibitemShut {NoStop}%
\bibitem [{\citenamefont {Evans}\ and\ \citenamefont
  {Morriss}(2008)}]{EvansNESM}%
  \BibitemOpen
  \bibfield  {author} {\bibinfo {author} {\bibfnamefont {D.~J.}\ \bibnamefont
  {Evans}}\ and\ \bibinfo {author} {\bibfnamefont {G.}~\bibnamefont
  {Morriss}},\ }\href@noop {} {\emph {\bibinfo {title} {Statistical Mechanics
  Of Nonequilibrium Liquids}}}\ (\bibinfo  {publisher} {Cambridge University
  Press},\ \bibinfo {year} {2008})\BibitemShut {NoStop}%
\bibitem [{\citenamefont {McQuarrie}(1976)}]{quarrie}%
  \BibitemOpen
  \bibfield  {author} {\bibinfo {author} {\bibfnamefont {D.~A.}\ \bibnamefont
  {McQuarrie}},\ }\href@noop {} {\emph {\bibinfo {title} {Statistical
  Mechanics}}}\ (\bibinfo  {publisher} {Harper and Row},\ \bibinfo {year}
  {1976})\BibitemShut {NoStop}%
\bibitem [{\citenamefont {Allen}(1984)}]{Allen84}%
  \BibitemOpen
  \bibfield  {author} {\bibinfo {author} {\bibfnamefont {M.~P.}\ \bibnamefont
  {Allen}},\ }\href@noop {} {\bibfield  {journal} {\bibinfo  {journal} {Mol.
  Phys.}\ }\textbf {\bibinfo {volume} {52}},\ \bibinfo {pages} {705} (\bibinfo
  {year} {1984})}\BibitemShut {NoStop}%
\bibitem [{\citenamefont {L{\'o}pez-Barr{\'o}n}, \citenamefont {Wagner},\ and\
  \citenamefont {Porcar}(2015)}]{Lopez2015}%
  \BibitemOpen
  \bibfield  {author} {\bibinfo {author} {\bibfnamefont {C.~R.}\ \bibnamefont
  {L{\'o}pez-Barr{\'o}n}}, \bibinfo {author} {\bibfnamefont {N.~J.}\
  \bibnamefont {Wagner}}, \ and\ \bibinfo {author} {\bibfnamefont
  {L.}~\bibnamefont {Porcar}},\ }\href@noop {} {\bibfield  {journal} {\bibinfo
  {journal} {J. Rheol.}\ }\textbf {\bibinfo {volume} {59}},\ \bibinfo {pages}
  {793} (\bibinfo {year} {2015})}\BibitemShut {NoStop}%
\bibitem [{\citenamefont {Lee}\ \emph {et~al.}(2018)\citenamefont {Lee},
  \citenamefont {Jiang}, \citenamefont {Wang}, \citenamefont {Sandy},\ and\
  \citenamefont {Lin}}]{Lee2018}%
  \BibitemOpen
  \bibfield  {author} {\bibinfo {author} {\bibfnamefont {J.}~\bibnamefont
  {Lee}}, \bibinfo {author} {\bibfnamefont {Z.}~\bibnamefont {Jiang}}, \bibinfo
  {author} {\bibfnamefont {J.}~\bibnamefont {Wang}}, \bibinfo {author}
  {\bibfnamefont {A.~R.}\ \bibnamefont {Sandy}}, \ and\ \bibinfo {author}
  {\bibfnamefont {S.~N. X.-M.}\ \bibnamefont {Lin}},\ }\href@noop {} {\bibfield
   {journal} {\bibinfo  {journal} {Phys. Rev. Lett.}\ }\textbf {\bibinfo
  {volume} {120}},\ \bibinfo {pages} {028002} (\bibinfo {year}
  {2018})}\BibitemShut {NoStop}%
\bibitem [{\citenamefont {Pan}\ \emph {et~al.}(2015)\citenamefont {Pan},
  \citenamefont {de~Cagny}, \citenamefont {Weber},\ and\ \citenamefont
  {Bonn}}]{Pan}%
  \BibitemOpen
  \bibfield  {author} {\bibinfo {author} {\bibfnamefont {Z.}~\bibnamefont
  {Pan}}, \bibinfo {author} {\bibfnamefont {H.}~\bibnamefont {de~Cagny}},
  \bibinfo {author} {\bibfnamefont {B.}~\bibnamefont {Weber}}, \ and\ \bibinfo
  {author} {\bibfnamefont {D.}~\bibnamefont {Bonn}},\ }\href@noop {} {\bibfield
   {journal} {\bibinfo  {journal} {Phys. Rev. E}\ }\textbf {\bibinfo {volume}
  {92}},\ \bibinfo {pages} {032202} (\bibinfo {year} {2015})}\BibitemShut
  {NoStop}%
\bibitem [{\citenamefont {Guy}, \citenamefont {Hermes},\ and\ \citenamefont
  {Poon}(2015)}]{Guy}%
  \BibitemOpen
  \bibfield  {author} {\bibinfo {author} {\bibfnamefont {B.~M.}\ \bibnamefont
  {Guy}}, \bibinfo {author} {\bibfnamefont {M.}~\bibnamefont {Hermes}}, \ and\
  \bibinfo {author} {\bibfnamefont {W.~C.~K.}\ \bibnamefont {Poon}},\
  }\href@noop {} {\bibfield  {journal} {\bibinfo  {journal} {Phys. Rev. Lett.}\
  }\textbf {\bibinfo {volume} {115}},\ \bibinfo {pages} {088304} (\bibinfo
  {year} {2015})}\BibitemShut {NoStop}%
\bibitem [{\citenamefont {Mari}\ \emph {et~al.}(2014)\citenamefont {Mari},
  \citenamefont {Seto}, \citenamefont {Morris},\ and\ \citenamefont
  {Denn}}]{Mari}%
  \BibitemOpen
  \bibfield  {author} {\bibinfo {author} {\bibfnamefont {R.}~\bibnamefont
  {Mari}}, \bibinfo {author} {\bibfnamefont {R.}~\bibnamefont {Seto}}, \bibinfo
  {author} {\bibfnamefont {F.~J.}\ \bibnamefont {Morris}}, \ and\ \bibinfo
  {author} {\bibfnamefont {M.~M.}\ \bibnamefont {Denn}},\ }\href@noop {}
  {\bibfield  {journal} {\bibinfo  {journal} {J. Rheol.}\ }\textbf {\bibinfo
  {volume} {58}},\ \bibinfo {pages} {1693} (\bibinfo {year}
  {2014})}\BibitemShut {NoStop}%
\bibitem [{\citenamefont {Heussinger}(2013)}]{Heussinger}%
  \BibitemOpen
  \bibfield  {author} {\bibinfo {author} {\bibfnamefont {C.}~\bibnamefont
  {Heussinger}},\ }\href@noop {} {\bibfield  {journal} {\bibinfo  {journal}
  {Phys. Rev. E}\ }\textbf {\bibinfo {volume} {88}},\ \bibinfo {pages} {050201}
  (\bibinfo {year} {2013})}\BibitemShut {NoStop}%
\bibitem [{\citenamefont {Seto}\ \emph
  {et~al.}(2013{\natexlab{b}})\citenamefont {Seto}, \citenamefont {Mari},
  \citenamefont {Morris},\ and\ \citenamefont {Denn}}]{Seto}%
  \BibitemOpen
  \bibfield  {author} {\bibinfo {author} {\bibfnamefont {R.}~\bibnamefont
  {Seto}}, \bibinfo {author} {\bibfnamefont {R.}~\bibnamefont {Mari}}, \bibinfo
  {author} {\bibfnamefont {F.~J.}\ \bibnamefont {Morris}}, \ and\ \bibinfo
  {author} {\bibfnamefont {M.~M.}\ \bibnamefont {Denn}},\ }\href@noop {}
  {\bibfield  {journal} {\bibinfo  {journal} {Phys. Rev. Lett.}\ }\textbf
  {\bibinfo {volume} {111}},\ \bibinfo {pages} {218301} (\bibinfo {year}
  {2013}{\natexlab{b}})}\BibitemShut {NoStop}%
\bibitem [{\citenamefont {Denn}, \citenamefont {Morris},\ and\ \citenamefont
  {Bonn}(2018)}]{Denn}%
  \BibitemOpen
  \bibfield  {author} {\bibinfo {author} {\bibfnamefont {M.~M.}\ \bibnamefont
  {Denn}}, \bibinfo {author} {\bibfnamefont {F.~J.}\ \bibnamefont {Morris}}, \
  and\ \bibinfo {author} {\bibfnamefont {D.}~\bibnamefont {Bonn}},\ }\href@noop
  {} {\bibfield  {journal} {\bibinfo  {journal} {Soft Matter}\ }\textbf
  {\bibinfo {volume} {14}},\ \bibinfo {pages} {170} (\bibinfo {year}
  {2018})}\BibitemShut {NoStop}%
\bibitem [{\citenamefont {Fernandez}\ \emph {et~al.}(2013)\citenamefont
  {Fernandez}, \citenamefont {Mani}, \citenamefont {Rinaldi}, \citenamefont
  {Kadau}, \citenamefont {Mosquet}, \citenamefont {Lombois-Burger},
  \citenamefont {Cayer-Barrioz}, \citenamefont {Herrmann}, \citenamefont
  {Spencer},\ and\ \citenamefont {Isa}}]{Fernandez2013}%
  \BibitemOpen
  \bibfield  {author} {\bibinfo {author} {\bibfnamefont {N.}~\bibnamefont
  {Fernandez}}, \bibinfo {author} {\bibfnamefont {R.}~\bibnamefont {Mani}},
  \bibinfo {author} {\bibfnamefont {D.}~\bibnamefont {Rinaldi}}, \bibinfo
  {author} {\bibfnamefont {D.}~\bibnamefont {Kadau}}, \bibinfo {author}
  {\bibfnamefont {M.}~\bibnamefont {Mosquet}}, \bibinfo {author} {\bibfnamefont
  {H.}~\bibnamefont {Lombois-Burger}}, \bibinfo {author} {\bibfnamefont
  {J.}~\bibnamefont {Cayer-Barrioz}}, \bibinfo {author} {\bibfnamefont {H.~J.}\
  \bibnamefont {Herrmann}}, \bibinfo {author} {\bibfnamefont {N.~D.}\
  \bibnamefont {Spencer}}, \ and\ \bibinfo {author} {\bibfnamefont
  {L.}~\bibnamefont {Isa}},\ }\href@noop {} {\bibfield  {journal} {\bibinfo
  {journal} {Phys. Rev. Lett.}\ }\textbf {\bibinfo {volume} {111}},\ \bibinfo
  {pages} {108301} (\bibinfo {year} {2013})}\BibitemShut {NoStop}%
\end{thebibliography}

\providecommand{\noopsort}[1]{}\providecommand{\singleletter}[1]{#1}%

\end{document}